\documentclass[11pt,oneolumn,floatfix,altaffilletter,superscriptaddress,tightenlines,showpacs,showkeys,preprintnumbers,nofootinbib]{revtex4-2}
\pdfoutput=1
\usepackage[colorlinks=true,citecolor=blue,linkcolor=blue,breaklinks=true,linktoc=none]{hyperref}
\usepackage{amsmath,amssymb}
\usepackage{epsfig} 
\usepackage{graphicx}
\usepackage{url}
\usepackage{color}
\usepackage{multirow}
\usepackage{placeins}
\usepackage{float}
\usepackage[dvipsnames]{xcolor}
 \usepackage{booktabs} 
\usepackage{braket}
\usepackage{float}
\usepackage{slashed,tabularray}
\hypersetup{
  colorlinks=true,
  linkcolor=red,
  filecolor=magenta,   
  urlcolor=blue,
  citecolor=blue,
}

\hypersetup{colorlinks,linkcolor={blue},citecolor={teal},urlcolor={violet}}

\allowdisplaybreaks

\bibpunct{[}{]}{,}{n}{}{,}

\def\beq{\begin{equation}}
\def\eeq{\end{equation}}
\def\bea{\begin{eqnarray}}
\def\eea{\end{eqnarray}}
\makeatletter
\@addtoreset{equation}{section}

\newcommand{\INFN}{INFN - Sezione di Napoli, Complesso Universitario Monte S. Angelo, I-80126 Napoli, Italy}

\newcommand{\SSM}{Scuola Superiore Meridionale, Università degli studi di Napoli ``Federico II'', Largo San Marcellino 10, 80138 Napoli, Italy}
\newcommand{\NAN}{Department of Physics and Institute of Theoretical Physics,
Nanjing Normal University, Nanjing, 210023, China}

\begin{document}

\title{PTA-Compatible Domain Walls at LISA and Taiji: Bayesian Reconstruction and Multiband Inference} 
\author{Satyabrata Datta}
    \email{amisatyabrata703@gmail.com}
    \affiliation{\NAN}
 \author{Rome Samanta}
  \email{samanta@na.infn.it}
  \affiliation{\SSM}
  \affiliation{\INFN}

\begin{abstract}
Domain walls provide an excellent fit to Pulsar Timing Array (PTA) data. A distinctive feature of the associated gravitational-wave (GW) spectrum is its ultraviolet (UV) tail, which, unlike that of many other cosmological GW sources, can extend into the mHz band and thereby render cross-detection with space-based interferometers such as LISA and Taiji possible. In this work, we explore the PTA-compatible parameter space of domain-wall models and investigate both the reconstruction prospects at LISA and Taiji and the information gain achievable through joint PTA--LISA and PTA--Taiji analyses. We find that LISA and Taiji can independently probe an extended region of parameter space that yields an ultraviolet tail in the milli-Hz band, even in the presence of astrophysical foregrounds. Within the PTA-supported region, however, the genuinely informative regime is more restricted and is concentrated toward the high-signal edge. In the space-based-detector-only analysis, the posterior is strongly degenerate in the underlying model parameters, although one principal parameter combination can be reconstructed with high precision where the signal is sufficiently strong. When PTA-informed priors are incorporated, the additional information gain is localized to the same high-signal region: there, the space-based data produce a non-negligible posterior update and efficiently suppress the catastrophic LISA(Taiji)-only degeneracy, while the final posterior often retains an axis ratio comparable to that of PTA alone. Our analysis is based on a 10-dimensional Bayesian inference, sampled with nested sampling, involving two domain-wall signal parameters and eight nuisance parameters describing instrumental noise and astrophysical foregrounds. For the LISA-only and Taiji-only analyses, we consider 49 injections distributed on an equidistant grid in the domain-wall parameter space, whereas for the joint analyses, 65 injections are selected from the PTA-supported region in order to quantify where the space-based data provide a meaningful posterior update beyond PTA alone. In both cases, we use Clough--Tocher interpolation to construct posterior heat maps over the full signal-parameter plane.
\end{abstract}
\maketitle
\tableofcontents

\section{Introduction}

The direct detection of gravitational waves (GWs) by LIGO opened a new observational window on the Universe and established GW astronomy as a precision probe of fundamental physics and astrophysics \cite{LIGOScientific:2016aoc}. Since then, the field has expanded rapidly across a broad range of frequencies. Ground-based interferometers probe the audio band, pulsar timing arrays (PTAs) access nanohertz frequencies, and space-based observatories such as LISA, Taiji, and TianQin are designed to explore the milli-Hz band \cite{LIGOScientific:2016jlg,ipta,lisa,taiji1,taiji2,TianQin}. In parallel, third-generation and decihertz concepts such as the Einstein Telescope (ET) and DECIGO promise to extend GW observations both in sensitivity and in frequency coverage \cite{et,decigo}. In this sense, GW astronomy is entering a genuinely multiband era, in which the same cosmological source class may be tested across many decades in frequency.

A major recent development in this direction has come from PTA experiments. The latest PTA datasets have reported evidence for a nanohertz GW background, or at least for a common-spectrum process with the expected inter-pulsar correlations, thereby providing the first compelling target for cosmological GW model building in the ultra-low-frequency regime \cite{epta,nanograv,ng1,ng2,ng3,ng4,ng5}. While the origin of the signal is not yet established, the PTA result has already changed the landscape of cosmological GW phenomenology: instead of asking only whether a given early-Universe source is observable in principle, one may now ask whether a source interpretation consistent with PTA data also predicts observable signatures at higher frequencies. PTA observations, therefore, define a phenomenologically motivated target region in parameter space against which future space-based and next-generation interferometers can be tested.

A wide variety of source classes have been proposed to account for the PTA signal, including supermassive black-hole binaries, phase transitions, inflationary scenarios, primordial black holes, and topological defects \cite{ng5}. Among cosmological sources, topological defects are especially interesting because they naturally generate broadband spectra extending over many decades in frequency. In particular, cosmic strings and domain walls can produce GW signals whose low-frequency behavior may be compatible with PTA observations while simultaneously yielding non-negligible power in the milli-Hz or deci-Hz bands \cite{ng5}. Such sources are therefore especially well suited for multiband tests, since they allow one to assess not only detectability at multiple instruments, but also the extent to which combining frequency bands sharpens parameter reconstruction that is intrinsically limited in any single-band analysis.

In this work, we focus on domain walls \cite{Zeldovich:1974uw,Press:1989yh,Saikawa:2017hiv}. A distinctive feature of the corresponding GW spectrum is that, unlike many cosmological backgrounds that are sharply localized in frequency, the domain-wall signal can possess a broad ultraviolet (UV) tail. As a result, a parameter region that fits PTA data at nanohertz frequencies can still leave a measurable imprint in the milli-Hz band accessible to LISA and Taiji \cite{ng5}. This makes domain walls a particularly attractive case study for multiband GW astronomy. At the same time, it also creates an inference problem of a distinctive kind: PTAs and space-based interferometers probe different parts of the same spectrum and therefore constrain different combinations of the underlying model parameters.

A realistic assessment of this possibility must account not only for instrumental noise, but also for the astrophysical foreground environment in the milli-Hz band. In particular, the unresolved Galactic population of compact white-dwarf binaries is expected to produce a confusion foreground that dominates a substantial part of the LISA band and constitutes one of the main obstacles to the extraction of weak cosmological backgrounds \cite{Adams:2013qma,Boileau:2020rpg,Boileau:2021sni,Korol:2021pun,Korol:2020lpq,Liu:2023qap}. In addition, extra-galactic compact-binary populations and other stochastic astrophysical contributions can further complicate the interpretation of a broad cosmological signal \cite{Phinney:2001di,Regimbau:2011rp,Babak:2023lro, Lehoucq:2023zlt}. Any claim of detectability or parameter reconstruction for a PTA-motivated cosmological source at LISA or Taiji must therefore be established in the presence of both instrumental noise and astrophysical foreground contamination.

These considerations raise several nontrivial questions. First, can PTA-compatible domain-wall signals be probed at mid-band interferometers such as LISA and Taiji once realistic instrumental noise and unresolved astrophysical foregrounds are included? Second, if the answer is yes, which combinations of the underlying signal parameters are actually constrained by the space-based data alone? Third, over what part of the PTA-supported parameter space does this single-detector constraint become genuinely informative, rather than prior-dominated? In what follows, we refer to this as the \emph{LISA-only} or \emph{Taiji-only} analysis, namely an inference based solely on the simulated space-based-detector dataset, without imposing PTA posterior information as an informative prior on the signal parameters. These questions are already nontrivial at the single-detector level, because a detector that probes only the UV tail need not determine all physical parameters independently. A narrow constraint on one effective parameter combination may coexist with a broad degeneracy in the original model parameter space, and within the PTA-supported region, this constraining power need not be uniform.

The multiband problem then adds two further questions. First, once PTA information is incorporated, in which part of the PTA-supported parameter space does LISA or Taiji provide a meaningful posterior update beyond PTA alone? Second, does the space-based likelihood primarily reduce the strong LISA(Taiji)-only degeneracy, or does it also substantially reduce the intrinsic elongation of the PTA posterior? By \emph{joint PTA--LISA} (or \emph{joint PTA--Taiji}) analysis, we mean an inference in which the PTA posterior information on the signal parameters is incorporated and then updated with the LISA (or Taiji) likelihood, so that both frequency bands contribute to the final posterior. These questions are conceptually distinct. A space-based detector may significantly sharpen the PTA-supported region and remove the catastrophic single-band LISA(Taiji) degeneracy, while still leaving the final posterior with an axis ratio comparable to that of PTA alone. In that case, the main multiband gain is not a dramatic reduction of the PTA elongation, but rather a localized tightening of the PTA-supported region in the high-signal part of parameter space.

A further key point of the present analysis is also methodological. In degenerate Bayesian inference problems, commonly used summaries such as one-dimensional marginal widths or total posterior area can be misleading. If the posterior develops a ridge, then the broad direction is often strongly prior sensitive and should not be interpreted as a clean measure of the constraining power of the data. Instead, the physically relevant quantity is the width across the ridge, namely the posterior variance perpendicular to the degenerate direction. This principal-component viewpoint provides a natural way to separate data-driven information from prior-dominated broad directions, and it also furnishes a convenient language for discussing the extent to which multiband data sharpen the posterior in the UV-tail regime.

Our analysis is based on a 10-dimensional Bayesian inference problem consisting of two domain-wall signal parameters and eight nuisance parameters describing instrumental noise and astrophysical foregrounds. For the LISA-only and Taiji-only analyses, we consider 49 injections distributed on an equidistant grid in the domain-wall signal parameter space and use Clough--Tocher interpolation \cite{LeeSchachter1980,RenkaCline1984,Nielson1983} to construct heat maps over the full signal-parameter plane \cite{ng5}. For the joint PTA--LISA and PTA--Taiji analyses, we instead use 65 irregularly placed injections within the PTA-supported region, since the objective is to quantify where the space-based data add non-negligible information once PTA constraints are already incorporated. The resulting inference outputs are interpolated with the same Clough--Tocher method to construct heat maps over the PTA-supported region. This setup cleanly separates the single-detector reconstruction problem from the corresponding multiband update problem and allows us to track how the inclusion of PTA information reshapes the posterior across the PTA-compatible region.

Our main findings may be summarized as follows. We find that LISA and Taiji can probe an extended region of the domain-wall parameter space that yields an ultraviolet tail in the milli-Hz band. Within the PTA-supported region, however, the genuinely informative single-detector regime is more restricted and is concentrated toward the high-SNR edge of PTA support (see, e.g., the bottom-left panel of Fig.~\ref{fig:fig1}). In the single-detector analysis, the posterior is strongly degenerate in the underlying model parameters, although one principal parameter combination can be reconstructed with high precision where the signal is sufficiently strong (see, e.g., the middle and bottom-left panels of Fig.~\ref{fig:fig1}). In the joint PTA--LISA(Taiji) analysis, the additional information gain is localized to the same high-signal region. There, the space-based likelihood produces a non-negligible posterior update and efficiently removes the catastrophic LISA(Taiji)-only degeneracy, while the final posterior often retains an axis ratio comparable to that of PTA alone (see, e.g., the right panel of Fig.~\ref{fig:fig7} and the top-right panel of Fig.~\ref{fig:fig8}). Domain walls therefore, provide a particularly clear example of the nuanced power of multiband GW inference: even when the added information is localized, and the relative elongation of the final posterior is not dramatically reduced, combining low- and high-frequency GW observations can still substantially shrink the allowed parameter region and sharpen the interpretation of the inferred source parameters.

The rest of this paper is organized as follows. In Sec.~\ref{sec:dw_gw} we summarize the domain-wall GW signal model and the origin of the high-frequency degeneracy relevant for space-based detectors. In Sec.~\ref{sec:setup_lisa_taiji} we describe the detector setup, astrophysical foregrounds, and Bayesian inference framework for LISA and Taiji. In Sec.~\ref{sec:pca_generalities} we introduce the principal-component characterization of posterior degeneracies and signal reconstruction. The LISA-only and Taiji-only analyses are presented in Sec.~\ref{sec:singleband}, while the joint PTA--LISA and PTA--Taiji analyses are discussed in Sec.~\ref{sec:joint}. We conclude in Sec.~\ref{sec:conclusion}.

\section{Stochastic gravitational waves from domain walls}
\label{sec:dw_gw}

Domain walls are two-dimensional topological defects that can form after the spontaneous breaking of a discrete symmetry, such as a $Z_2$ symmetry \cite{Zeldovich:1974uw,Press:1989yh,Saikawa:2017hiv}. After their formation, domain-wall networks are expected to rapidly approach a scaling regime, in which the so-called area parameter $\mathcal{A}$ becomes approximately constant and of order unity, as found in lattice simulations \cite{Press:1989yh,Hiramatsu:2013qaa,Dankovsky:2024zvs,Babichev:2025stm,Notari:2025kqq,Cyr:2025nzf}. In this regime, the energy density stored in the network can be written as
\begin{equation}
    \rho_{\rm DW}\simeq \mathcal{A} H \sigma,
\end{equation}
where $H$ is the Hubble parameter and $\sigma$ is the domain-wall tension.

Since the total background energy density scales as $H^2$, the fractional contribution of domain walls to the cosmic energy budget grows as $H^{-1}$. Consequently, if the walls were perfectly stable, they would eventually dominate the energy density of the Universe, which is incompatible with standard cosmology. In simple symmetry realizations such as $Z_2$, the standard way to avoid this problem is to introduce a small bias term in the potential, which lifts the degeneracy between the vacua and causes the walls to annihilate at a characteristic temperature $T_*$. Between formation and annihilation, the wall network sources a sizable anisotropic stress and thereby emits a stochastic gravitational-wave background. The GW production is dominated near the annihilation epoch, and the characteristic emission scale is set by the Hubble scale at that time, $H_*$.

Since our goal is to study the PTA-motivated region of parameter space, we adopt the phenomenological template used in PTA analyses for the present-day GW energy density from domain walls \cite{ng5},
\begin{equation}
    \Omega_{\rm GW} h^2
    =
    \frac{3}{32\pi}\,
    \mathcal{D}\,
    \tilde{\epsilon}\,
    \alpha_*^2\,
    \mathcal{S}(f/f_p),
    \label{eq:omega_dw}
\end{equation}
where $\mathcal{D}\simeq 2\times 10^{-5}$ is a dilution factor, $\tilde{\epsilon}\simeq 0.7$ is an efficiency factor \cite{Hiramatsu:2013qaa}, and $\alpha_*$ denotes the fraction of the total energy density stored in domain walls at the annihilation temperature $T_*$. The function $\mathcal{S}$ describes the spectral shape and is obtained by fitting lattice-generated spectra with the form
\begin{equation}
    \mathcal{S}(x)=
    \frac{(a+b)^c}
    {\left(b\,x^{-a/c}+a\,x^{b/c}\right)^c},
    \qquad x \equiv \frac{f}{f_p}.
    \label{eq:dw_shape}
\end{equation}
This parametrization interpolates between an infrared power-law rise and an ultraviolet power-law tail, with the precise slopes determined by the parameters $a$, $b$, and $c$.

The peak frequency redshifted to the present epoch is given by
\begin{equation}
    f_p =
    1.14\,{\rm nHz}\,
    \left(\frac{10.75}{g_{*,s}}\right)^{1/3}
    \left(\frac{g_*}{10.75}\right)^{1/2}
    \left(\frac{T_*}{10\,{\rm MeV}}\right),
    \label{eq:fp_dw}
\end{equation}
where $g_*$ and $g_{*,s}$ are the effective relativistic degrees of freedom for the energy density and entropy density, respectively, evaluated at $T_*$.

In the main text, we adopt
\begin{equation}
    a=3,\qquad b=1,\qquad c=1,
\end{equation}
which may be regarded as a representative average over several simulation outputs and yields the familiar broken-power-law behavior with an infrared rise $\propto f^3$ and an ultraviolet decay $\propto f^{-1}$. In the Appendix, we also consider a spectral fit corresponding to a specific simulation output \cite{Babichev:2025stm}, with parameters slightly different from the benchmark choice above, and show that our main conclusions are not significantly affected.

For the discussion below, it is useful to make explicit the high-frequency limit of the spectrum. In the regime
\begin{equation}
    f \gg f_p,
\end{equation}
Eq.~\eqref{eq:dw_shape} simplifies to
\begin{equation}
    \mathcal{S}(f/f_p) \simeq
    \frac{(a+b)^c}{a^c}
    \left(\frac{f}{f_p}\right)^{-b}.
\end{equation}
For the benchmark choice $a=3$, $b=1$, $c=1$, this becomes
\begin{equation}
    \mathcal{S}(f/f_p)\simeq \frac{4}{3}\left(\frac{f}{f_p}\right)^{-1}.
\end{equation}
Substituting Eq.~\eqref{eq:fp_dw} into Eq.~\eqref{eq:omega_dw}, one finds
\begin{equation}
    \Omega_{\rm GW} h^2
    \propto
    \alpha_*^2 f_p
    \propto
    \alpha_*^2 T_*,
    \qquad f\gg f_p,
    \label{eq:uv_scaling}
\end{equation}
up to factors that depend only weakly on the thermal history through $g_*$ and $g_{*,s}$, and apart from the explicit overall frequency dependence $\propto f^{-1}$.

Equation~\eqref{eq:uv_scaling} is central to the inference problem at LISA and Taiji. For PTA-motivated domain-wall signals, the peak frequency typically lies far below the milli-Hz band, so space-based interferometers probe only the ultraviolet tail. In this regime, the signal depends predominantly on the combination $\alpha_*^2 T_*$ rather than on $\alpha_*$ and $T_*$ separately. As a result, a LISA-only or Taiji-only analysis generically develops a strong parameter degeneracy: many pairs $(\alpha_*,T_*)$ that satisfy approximately
\begin{equation}
    \alpha_*^2 T_* \simeq \mathrm{const.}
\end{equation}
produce nearly indistinguishable in-band spectra. In logarithmic variables, this corresponds to
\begin{equation}
    2\log_{10}\alpha_* + \log_{10}T_* \simeq \mathrm{const.},
\end{equation}
which explains the elongated anti-correlated posterior ridge encountered in the space-based-detector-only analysis. The data can therefore constrain one principal parameter combination quite well, while leaving the orthogonal combination only weakly determined.

This high-frequency simplification will play an important role throughout the analysis. It clarifies both why the LISA/Taiji-only posterior is strongly degenerate and why a principal-component characterization of the posterior geometry is the appropriate language in which to quantify the constrained and unconstrained directions in parameter space.

\section{General setup for parameter inference with LISA and Taiji}
\label{sec:setup_lisa_taiji}

In this section, we summarize the detector and statistical framework used in our analysis, including the construction of the likelihood function and the generation of mock data realizations. We begin with time-delay interferometry (TDI), a data-processing technique that suppresses the otherwise overwhelming laser-frequency noise and yields the orthogonal $A$, $E$, and $T$ data channels used in stochastic-background searches \cite{Tinto:2004wu,McNamara:2008zz,Armano:2018kix,Caprini:2019pxz,Flauger:2020qyi,Caprini:2024hue}. We then describe the frequency-domain modeling of the detector response, instrumental noise, and astrophysical foregrounds. Here and throughout, PSD denotes the one-sided power spectral density. Finally, we introduce the Bayesian likelihood and model-selection framework used in the parameter inference.

The analysis pipeline for LISA and Taiji is the same. The difference between the two detectors enters only through detector-specific instrumental parameters, most notably the arm length $L$ and the optical metrology noise level $\delta x$. These quantities determine the transfer frequency $f_\star$, the response functions, and the instrumental-noise PSDs, and hence the overall sensitivity. Synthetic datasets are generated using the corresponding detector-dependent spectra, while the subsequent Bayesian analysis proceeds with the same likelihood formalism in both cases.

\subsection{TDI channels, detector response, and instrumental noise}

LISA consists of a triangular constellation of three spacecraft separated by arm lengths
\begin{equation}
    L = 2.5\times 10^9~{\rm m},
\end{equation}
while for Taiji we take
\begin{equation}
    L = 3\times 10^9~{\rm m}.
\end{equation}
Each spacecraft exchanges laser beams with the other two, forming unequal-arm Michelson-type interferometric measurements. These raw measurements are linearly recombined by TDI into the orthogonal channels
\begin{equation}
    \tilde d_j(f),\qquad j\in\{A,E,T\}.
\end{equation}

For each TDI channel, the one-sided PSD is written as
\begin{equation}
    P_a(f)=S_a(f)+N_a(f),
    \qquad a\in\{A,E,T\},
\end{equation}
where $S_a(f)$ is the stochastic GW signal contribution and $N_a(f)$ is the instrumental-noise contribution. The signal term is related to the GW energy-density spectrum through
\begin{equation}
    S_a(f)
    =
    \frac{3H_0^2}{4\pi^2}\,
    \frac{\Omega_{\rm GW}(f)}{f^3}\,
    \mathcal{R}_a(f),
\end{equation}
where $\mathcal{R}_a(f)$ is the detector response function.

 The  response functions of the $A$, $E$, and $T$ channels are~\cite{Cornish:2001bb,Smith:2019wny}
\begin{equation}
    \mathcal{R}_A(f)=\mathcal{R}_E(f)
    =
    \frac{9}{5}|W(f)|^2
    \left[
    1+\left(\frac{f}{4f_\star/3}\right)^2
    \right]^{-1},
\end{equation}
\begin{equation}
    \mathcal{R}_T(f)
    =
    \frac{1}{1008}|W(f)|^2
    \left(\frac{f}{f_\star}\right)^6
    \left[
    1+\frac{5}{16128}\left(\frac{f}{f_\star}\right)^8
    \right]^{-1},
\end{equation}
with
\begin{equation}
    W(f)=1-e^{-2if/f_\star},
    \qquad
    f_\star=\frac{c}{2\pi L}.
\end{equation}
We adopt the usual $\Delta L/L$ convention, where $\Delta L$ is the difference in optical path lengths between the two interferometer arms and $L$ is the single-arm length.

The dominant raw instrumental contamination is laser-frequency noise, which is many orders of magnitude larger than the GW signal. TDI suppresses this contribution by constructing suitable combinations of delayed measurements \cite{Tinto:2004wu}. In the equal-arm approximation adopted here, this cancellation is encoded in the transfer factor $W(f)$. At low frequency, the $T$ channel has a much weaker response to GWs than the $A$ and $E$ channels, and therefore acts approximately as a null channel. In stochastic-background searches, this makes the $T$ channel particularly useful for monitoring instrumental noise.

The instrumental noise model is specified by two primary ingredients: the acceleration noise and the optical metrology system noise. The acceleration noise is taken as
\begin{equation}
    \sqrt{S_{\rm acc}(f)}
    =
    N_{\rm acc}
    \sqrt{1+\left(\frac{0.4~{\rm mHz}}{f}\right)^2}
    \sqrt{1+\left(\frac{f}{8~{\rm mHz}}\right)^4}
    \left(
    \frac{\rm m}{{\rm s}^2\sqrt{\rm Hz}}
    \right),
\end{equation}
with
\begin{equation}
    N_{\rm acc}=3\times 10^{-15}.
\end{equation}
The optical metrology noise is modeled as
\begin{equation}
    \sqrt{S_{\rm OMS}(f)}
    =
    \delta x
    \sqrt{1+\left(\frac{2~{\rm mHz}}{f}\right)^4}
    \left(
    \frac{\rm m}{\sqrt{\rm Hz}}
    \right).
\end{equation}
For LISA, we use
\begin{equation}
    \delta x = 15\times 10^{-12},
\end{equation}
whereas for Taiji, we adopt
\begin{equation}
    \delta x = 8\times 10^{-12}.
\end{equation}
Thus, when switching from LISA to Taiji in the analysis, the quantities that must be changed are
\begin{equation}
    L,\qquad \delta x,\qquad f_\star=\frac{c}{2\pi L},
\end{equation}
and consequently all response functions and noise PSDs that depend on them.

Before diagonalization into the $A$, $E$, and $T$ channels, the instrumental-noise combinations are
\begin{equation}
    N_A = N_E = N_1-N_2,
    \qquad
    N_T = N_1+2N_2,
\end{equation}
with
\begin{equation}
    N_1(f)
    =
    \frac{1}{L^2}
    \left\{
    4 S_{\rm OMS}(f)
    +
    8\left[
    1+\cos^2\left(\frac{f}{f_\star}\right)
    \right]
    \frac{S_{\rm acc}(f)}{(2\pi f)^4}
    \right\}
    |W(f)|^2,
\end{equation}
\begin{equation}
    N_2(f)
    =
    -\frac{1}{L^2}
    \left[
    2S_{\rm OMS}(f)
    +
    \frac{8S_{\rm acc}(f)}{(2\pi f)^4}
    \right]
    \cos\left(\frac{f}{f_\star}\right)
    |W(f)|^2.
\end{equation}

The effective sensitivity function of each TDI channel is defined by \cite{Babak:2021mhe}
\begin{equation}
    \mathcal{S}_j(f)=\sqrt{\frac{N_j(f)}{\mathcal{R}_j(f)}},
    \qquad j\in\{A,E,T\}.
\end{equation}
The corresponding channel-by-channel signal-to-noise ratio is
\begin{equation}
    {\rm SNR}_j
    =
    \left[
    2T_t\int_0^\infty
    \left(\frac{S_j(f)}{N_j(f)}\right)^2
    df
    \right]^{1/2},
    \qquad
    j\in\{A,E,T\},
\end{equation}
where $T_t$ is the total observation time. In practice, the integral is restricted to the analyzed detector band. Since the null-channel method assumes that the $T$ channel is effectively signal-free, the $A$ and $E$ channels carry the main signal sensitivity.

\subsection{Astrophysical foregrounds and Bayesian model comparison}

In the frequency band relevant for LISA and Taiji, a cosmological stochastic background is not observed in isolation. Rather, it is superimposed on astrophysical foregrounds and backgrounds whose origin is by now reasonably well understood from compact-binary population synthesis and recent LISA forecasting studies. The dominant contribution in the milli-Hz band is expected to come from the Galactic population of double white-dwarf (DWD) binaries, namely binary systems composed of two white dwarfs \cite{Korol:2021pun,Korol:2020lpq,Liu:2023qap}. Because the Milky Way contains a very large population of such systems, only the loudest subset will be individually resolvable, while the superposition of the unresolved binaries forms a confusion foreground. This Galactic foreground is not merely an effective noise term: it is a genuine astrophysical GW signal, expected to be prominent in LISA-like detectors and to carry information about Galactic structure and binary evolution. At the same time, for searches of weak cosmological backgrounds, it acts as the leading source of confusion in a substantial part of the milli-Hz band. 

In addition to the Galactic DWD foreground, unresolved extra-galactic compact binaries generate a smoother astrophysical stochastic background. In the current state of the art, this component is usually modeled in an effective way, since its detailed shape depends on assumptions about cosmic star-formation history, metallicity evolution, delay-time distributions, and binary evolution channels. Recent population studies indicate that extra-galactic white-dwarf binaries can themselves provide an important stochastic contribution in the LISA band, while stellar-origin black-hole and neutron-star binaries can also contribute, especially away from the peak of the Galactic confusion foreground. For the purposes of cosmological searches, however, these extra-galactic contributions are often well approximated by a smooth power-law or broken-power-law component over the limited frequency range of interest. 

Motivated by this picture, we model the total GW spectrum as
\begin{equation}
    \Omega_{\rm GW}(f)
    =
    \Omega_{\rm dwd}(f)
    +
    \Omega_{\rm GW,ast}(f)
    +
    \Omega_{\rm BSM}(f),
\end{equation}
where the three terms denote, respectively, the Galactic DWD foreground, the smoother extra-galactic astrophysical background, and the beyond-the-Standard-Model (BSM) signal of interest.

For the Galactic foreground we adopt the phenomenological broken-power-law form \cite{Korol:2021pun,Korol:2020lpq,Liu:2023qap,Chen:2023zkb}
\begin{equation}
    \Omega_{\rm dwd}(f)
    =
    \frac{A_1 (f/f_\star)^{\alpha_1}}
    {1+A_2 (f/f_\star)^{\alpha_2}}.
\end{equation}
This parametrization is not meant to capture every detail of the underlying Galactic population, such as anisotropy or the time modulation induced by the detector’s orbital motion \cite{Giampieri:1997ie,Criswell:2024hfn,Digman:2022jmp,Hindmarsh:2024ttn,Buscicchio:2024wwm}, but rather to provide an effective description of the broad spectral rise and turnover associated with the unresolved DWD confusion signal. Such phenomenological descriptions are widely used in forecasting studies, where the main goal is to assess how a cosmological signal competes spectrally with the dominant Galactic foreground. 

The extra-galactic astrophysical background is modeled as a single power law \cite{Phinney:2001di,Regimbau:2011rp,Babak:2023lro, Lehoucq:2023zlt,Boileau:2020rpg},
\begin{equation}
    \Omega_{\rm GW,ast}(f)
    =
    \Omega_{\rm ast}
    \left(\frac{f}{f_\star}\right)^{\varepsilon},
\end{equation}
where $\Omega_{\rm ast}$ is the amplitude at the reference frequency and $\varepsilon$ is the spectral index. This effective description should be understood as a proxy for the aggregate contribution from unresolved extra-galactic compact binaries. While more detailed population models can certainly be considered, the power-law approximation is sufficient for the present purpose, namely to quantify whether the domain-wall signal remains identifiable once the dominant astrophysical components are included.

The full set of free parameters used in the inference is listed in Table~\ref{tab:model_parameters}. These include the instrumental-noise parameters, the Galactic DWD foreground parameters, the astrophysical background parameters, and the two signal parameters of the domain-wall model. In the Taiji analysis, the same parameterization is adopted, with the detector-specific fiducial values modified as discussed above.

\begin{table}[htbp]
\centering
\begin{tblr}
	{
    hlines,
    vlines,
    row{1} = {bg=gray7, fg=white, font=\bfseries},
column{1} = {bg=gray9},
cell{1}{1} = {bg=gray7, fg=white},
}

\textbf{Parameter} & \textbf{Injected value} & \textbf{Prior (uniform)} \\
$\log_{10}(N_{\rm acc})$      & $-14.523$ & $(-16.0,-13.7)$ \\
$\log_{10}(\delta_x)$         & $-11.097$ & $(-13.0,-10.7)$ \\
$\log_{10}(A_1)$              & $-15.4$   & $(-18.0,-5.0)$ \\
$\alpha_1$                    & $-5.7$    & $(-15.0,-3.0)$ \\
$\log_{10}(A_2)$              & $-6.32$   & $(-10.0,5.0)$ \\
$\alpha_2$                    & $-6.2$    & $(-10.0,-1.0)$ \\
$\log_{10}(\Omega_{\rm ast})$ & $-11.0$   & $(-15.0,-8.0)$ \\
$\varepsilon$                 & $0.67$    & $(0.0,1.0)$ \\
$\log_{10}(T_\star)$          & grid      & $(-2.5,2.5)$ \\
$\log_{10}(\alpha_\star)$     & grid      & $(-2.5,0.5)$ \\
\end{tblr}
\caption{Model parameters, fiducial values, and uniform prior ranges used in the inference.}
\label{tab:model_parameters}
\end{table}

Although the present domain-wall signal is not strongly degenerate with the astrophysical components over most of the parameter space considered here, it is still useful to supplement the reconstruction analysis with a Bayesian model-comparison diagnostic. The reason is that signal visibility above instrumental noise does not automatically imply that the data statistically require an additional BSM component once astrophysical foreground freedom is included. In particular, one may already find a well-constrained principal signal combination while the overall improvement in fit remains insufficient to compensate for the Occam penalty associated with the enlarged parameter space of the extended model. For this reason, Bayesian model selection is included here as a secondary, detection-oriented diagnostic rather than as the main focus of the analysis.

For a model with parameters $\theta$, likelihood $\mathcal{L}(\theta)$, and prior $\pi(\theta)$, the Bayesian evidence is \cite{Trotta:2008qt, Thrane_2019}
\begin{equation}
    Z=\int \mathcal{L}(\theta)\,\pi(\theta)\,d\theta.
\end{equation}
We compare an astrophysical-only model with a model containing both astrophysical components and the BSM signal. The corresponding Bayes factor is
\begin{equation}
    {\rm BF}
    =
    \frac{Z_{\rm BSM+astro}}{Z_{\rm astro}},
\end{equation}
and we use $\log_{10}{\rm BF}$ as a diagnostic of model preference. In particular, values such as
\begin{equation}
    \log_{10}{\rm BF} > 1
\end{equation}
correspond to ${\rm BF}>10$, indicating that the data favor the extended model over the astrophysical-only hypothesis. The Bayes factor therefore, provides a useful complement to the instrumental-noise-based SNR, especially in the low-SNR regime where signal visibility and statistical preference need not coincide \cite{Kass:1995loi, Trotta:2008qt}.
\subsection{The frequency-domain likelihood}
\label{sec:likelihood_derivation}

We now derive the frequency-domain likelihood used in our Bayesian analysis and summarize the procedure employed to generate mock data realizations. The key idea is that the TDI data in each frequency bin are modeled as complex Gaussian random variables whose covariance is determined by the total power spectral density, i.e.\ the sum of signal and noise contributions.

Let $d_a(t)$ denote the time-domain data in the TDI channel $a\in\{A,E,T\}$. The total observation is divided into $N_0$ valid time segments of duration $T$, labeled by $\kappa=1,\dots,N_0$. Within each segment, the data are sampled at intervals $\Delta t$, so that the sampling frequency is
$
f_s=\Delta t^{-1},
$
and the number of time samples per segment is
$
N=T/\Delta t.
$
We assume that each segment is sufficiently short that the data can be treated as stationary within that segment.

The discrete Fourier transform (DFT) of the $\kappa$-th segment is written as
\begin{equation}
    \tilde d_a^\kappa(f_k)
    =
    \sum_{n=0}^{N-1}
    d_a^\kappa(t_n)\,
    e^{-2\pi i kn/N},
\end{equation}
where the discrete frequencies are
\begin{equation}
    f_k = k \Delta f,
    \qquad
    \Delta f = \frac{1}{T},
    \qquad
    k=0,1,\dots,\frac{N}{2}.
\end{equation}
Each time segment therefore provides data across the analyzed frequency band, from the lowest nonzero Fourier bin $f_{\min}=1/T$ up to the Nyquist frequency $f_{\max}=f_s/2$. For each segment $\kappa$ and frequency bin $f_k$, we define the three-component complex data vector
\begin{equation}
    \mathbf d_{\kappa k}
    \equiv
    \begin{pmatrix}
        \tilde d_A^\kappa(f_k)\\[3pt]
        \tilde d_E^\kappa(f_k)\\[3pt]
        \tilde d_T^\kappa(f_k)
    \end{pmatrix}.
\end{equation}

The standard SGWB likelihood assumes that, for each segment and each frequency bin, the Fourier coefficients are drawn from a zero-mean complex Gaussian distribution. In our simplified treatment, the $A$, $E$, and $T$ channels are taken to be statistically independent in the frequency domain, so that the covariance matrix is diagonal. Let the total one-sided PSD in channel $a$ be denoted by
\begin{equation}
    P_a(f;\theta)=S_a(f;\theta)+N_a(f;\theta),
    \qquad a\in\{A,E,T\}.
\end{equation}
In the null-channel approximation, the $T$ channel is assumed to contain a negligible GW signal, so that
\begin{equation}
    P_A(f;\theta)=S_A(f;\theta)+N_A(f;\theta),\qquad
    P_E(f;\theta)=S_E(f;\theta)+N_E(f;\theta),\qquad
    P_T(f;\theta)=N_T(f;\theta).
\end{equation}
The signal in each TDI channel is related to the underlying GW energy-density spectrum through
\begin{equation}
    S_a(f;\theta)
    =
    \frac{3H_0^2}{4\pi^2}\,
    \frac{\Omega_{\rm GW}(f;\theta)}{f^3}\,
    \mathcal R_a(f),
\end{equation}
so the model parameters $\theta$ enter the likelihood through the predicted spectra $P_a(f;\theta)$.

With our DFT convention, the covariance of one complex Fourier coefficient is \cite{Pieroni:2020rob}
\begin{equation}
    \big\langle |\tilde d_a^\kappa(f_k)|^2 \big\rangle
    =
    \frac{T f_s^2}{2}\,P_a(f_k;\theta).
\end{equation}
Thus the covariance matrix for $\mathbf d_{\kappa k}$ is
\begin{equation}
    C_k(\theta)
    =
    \frac{T f_s^2}{2}
    \begin{pmatrix}
        P_A(f_k;\theta) & 0 & 0\\
        0 & P_E(f_k;\theta) & 0\\
        0 & 0 & P_T(f_k;\theta)
    \end{pmatrix}.
    \label{eq:covariance_matrix}
\end{equation}

For an $n$-dimensional zero-mean complex Gaussian vector $\mathbf d$ with covariance matrix $C$, the probability density is
\begin{equation}
    p(\mathbf d|\theta)
    =
    \frac{1}{\pi^n \det C}
    \exp\!\left(-\mathbf d^\dagger C^{-1}\mathbf d\right).
\end{equation}
In the present case $n=3$, so for one segment $\kappa$ and one frequency bin $f_k$ we have
\begin{equation}
    p(\mathbf d_{\kappa k}|\theta)
    =
    \frac{1}{\pi^3 \det C_k(\theta)}
    \exp\!\left[-\mathbf d_{\kappa k}^\dagger C_k^{-1}(\theta)\mathbf d_{\kappa k}\right].
\end{equation}
Using Eq.~\eqref{eq:covariance_matrix}, the determinant is
\begin{equation}
    \det C_k(\theta)
    =
    \left(\frac{T f_s^2}{2}\right)^3
    P_A(f_k;\theta)\,P_E(f_k;\theta)\,P_T(f_k;\theta),
\end{equation}
and the quadratic form becomes
\begin{equation}
    \mathbf d_{\kappa k}^\dagger C_k^{-1}(\theta)\mathbf d_{\kappa k}
    =
    \frac{2}{T f_s^2}
    \left[
        \frac{|\tilde d_A^\kappa(f_k)|^2}{P_A(f_k;\theta)}
        +
        \frac{|\tilde d_E^\kappa(f_k)|^2}{P_E(f_k;\theta)}
        +
        \frac{|\tilde d_T^\kappa(f_k)|^2}{P_T(f_k;\theta)}
    \right].
\end{equation}

Assuming statistical independence between different segments and different Fourier bins, the full likelihood is obtained by multiplying the one-bin likelihood over all $\kappa$ and $k$,
\begin{equation}
    \mathcal L(\theta)
    =
    \prod_{\kappa=1}^{N_0}
    \prod_{k=1}^{N/2}
    p(\mathbf d_{\kappa k}|\theta).
\end{equation}
Taking the logarithm and dropping additive constants independent of $\theta$, one finds
\begin{equation}
\begin{aligned}
    \ln \mathcal L(\theta)
    =
    -\sum_{\kappa=1}^{N_0}\sum_{k=1}^{N/2}
    \Bigg[
        &\ln\!\Big(P_A(f_k;\theta)\,P_E(f_k;\theta)\,P_T(f_k;\theta)\Big) \\
        &+
        \frac{2}{T f_s^2}
        \left(
        \frac{|\tilde d_A^\kappa(f_k)|^2}{P_A(f_k;\theta)}
        +
        \frac{|\tilde d_E^\kappa(f_k)|^2}{P_E(f_k;\theta)}
        +
        \frac{|\tilde d_T^\kappa(f_k)|^2}{P_T(f_k;\theta)}
        \right)
    \Bigg]
    +{\rm const.}
\end{aligned}
\label{eq:loglike_compact}
\end{equation}
or, using the null-channel approximation explicitly,
\begin{equation}
\begin{aligned}
    \ln \mathcal L(\theta)
    =
    -\sum_{\kappa=1}^{N_0}\sum_{k=1}^{N/2}
    \Bigg[
        &\ln\!\Big((S_A(f_k;\theta)+N_A(f_k;\theta))
        (S_E(f_k;\theta)+N_E(f_k;\theta))
        N_T(f_k;\theta)\Big) \\
        &+
        \frac{2}{T f_s^2}
        \left(
        \frac{|\tilde d_A^\kappa(f_k)|^2}{S_A(f_k;\theta)+N_A(f_k;\theta)}
        +
        \frac{|\tilde d_E^\kappa(f_k)|^2}{S_E(f_k;\theta)+N_E(f_k;\theta)}
        +
        \frac{|\tilde d_T^\kappa(f_k)|^2}{N_T(f_k;\theta)}
        \right)
    \Bigg]
    +{\rm const.}
\end{aligned}
\label{eq:loglike_full}
\end{equation}
which is the form implemented in the numerical analysis. The current form of $\mathcal{L}(\theta)$ has also been used in Ref.~\cite{Guan:2025idx}. In the literature, an alternative Gaussian-residual likelihood has also been considered; see, for example, Refs.~\cite{Caprini:2019pxz,Flauger:2020qyi,Samanta:2025jec}. The two likelihoods are based on different statistical assumptions about the frequency-domain estimator, but in the limit of a large number of averaged segments, the estimator becomes approximately Gaussian, so the two approaches typically yield very similar parameter-inference results.

For forecasts and injection studies, synthetic data realizations are generated from a chosen fiducial model $\theta_{\rm inj}$. For each segment $\kappa$ and frequency bin $f_k$, one draws the mock data vector
$
\mathbf d_{\kappa k}^{\rm mock}
$
from a zero-mean complex Gaussian with covariance $C_k^{\rm inj}$,
\begin{equation}
    \mathbf d_{\kappa k}^{\rm mock}
    \sim
    \mathcal N_{\mathbb C}\!\left(0,C_k^{\rm inj}\right).
\end{equation}
Since the covariance is diagonal, this is equivalent to drawing each channel independently:
\begin{equation}
    \tilde d_a^\kappa(f_k)
    =
    \sqrt{\frac{T f_s^2}{4}\,P_a^{\rm inj}(f_k)}
    \left(x_{a,\kappa k}+i\,y_{a,\kappa k}\right),
\end{equation}
where $x_{a,\kappa k}$ and $y_{a,\kappa k}$ are independent standard normal random variables,
$
x_{a,\kappa k},y_{a,\kappa k}\sim\mathcal N(0,1).
$
This normalization guarantees that
\begin{equation}
    \big\langle |\tilde d_a^\kappa(f_k)|^2 \big\rangle
    =
    \frac{T f_s^2}{2}\,P_a^{\rm inj}(f_k),
\end{equation}
as required. Repeating this draw for all
$
\kappa=1,\dots,N_0
$
and
$
k=1,\dots,N/2
$
produces a complete mock dataset
\begin{equation}
    \left\{
    \tilde d_A^\kappa(f_k),\,
    \tilde d_E^\kappa(f_k),\,
    \tilde d_T^\kappa(f_k)
    \right\},
\end{equation}
which is then analyzed with the likelihood of Eq.~\eqref{eq:loglike_full}.

With this likelihood in hand, the Bayesian analysis proceeds in the standard way. Denoting the full parameter vector by $\theta$, the posterior distribution of the ten model parameters is
\begin{equation}
    p(\theta|d)
    =
    \frac{\mathcal L(d|\theta)\,\pi(\theta)}{Z},
\end{equation}
where $\pi(\theta)$ is the prior distribution and
\begin{equation}
    Z=\int d\theta\, \mathcal L(d|\theta)\,\pi(\theta)
\end{equation}
is the Bayesian evidence. In practice, we sample this posterior numerically to obtain the multidimensional posterior distribution and its various marginalizations for the ten parameters of interest. Parameter-space exploration is performed using the \texttt{dynesty} sampler \cite{speagle2020dynesty}, interfaced through the \texttt{Bilby} package \cite{Ashton:2018jfp}.

\section{Principal-component analysis of posterior degeneracies and signal reconstruction}
\label{sec:pca_generalities}

In the LISA-only and Taiji-only analyses considered here, the central issue is not whether the underlying physical parameters can be recovered individually, but rather which \emph{combinations} of parameters are actually constrained by the data. This distinction becomes essential when the signal enters the detector band only through its ultraviolet tail, so that many parameter choices generate nearly indistinguishable in-band spectra. In such cases, the posterior may be sharply confined across one direction in parameter space while remaining broad along another. Consequently, one-dimensional marginal widths or posterior areas can give a misleading impression of the true constraining power of the experiment. A more informative characterization is provided by a principal-component analysis (PCA) of the posterior covariance \cite{PCA}. Unlike the Fisher information matrix, which relies on a local quadratic approximation of the likelihood and can be ill-conditioned in the presence of strong degeneracies \cite{Vallisneri:2007ev}, a PCA of the posterior samples captures the global topology of the parameter space.

Let the model be parametrized by an $N$-dimensional vector
\[
\boldsymbol{\theta}=(\theta_1,\theta_2,\dots,\theta_N)^T,
\]
with posterior distribution $p(\boldsymbol{\theta}\mid d)$. Even when the posterior is not exactly Gaussian, its sample covariance matrix
\[
C\equiv {\rm Cov}(\boldsymbol{\theta})
\]
provides a useful second-moment summary of its local geometry. Since $C$ is real and symmetric, it admits the eigen decomposition
\[
C=E\Lambda E^T,
\]
where the columns of $E$ are orthonormal eigenvectors $\mathbf e_i$ and
\[
\Lambda={\rm diag}(\lambda_1,\lambda_2,\dots,\lambda_N),
\qquad
\lambda_1\ge\lambda_2\ge\cdots\ge\lambda_N\ge 0.
\]
The eigenvectors define the principal directions of the posterior cloud, while the eigenvalues give the variances along those directions. A strong hierarchy among the $\lambda_i$ therefore signals a parameter degeneracy.

For the present problem, the most relevant case is the two-parameter posterior in the logarithmic variables
\[
\theta_1=\log_{10}\alpha_*,
\qquad
\theta_2=\log_{10}T_*,
\]
for which multiplicative degeneracies are represented approximately linearly. The covariance matrix can then be written as
\begin{equation}
C=
\begin{pmatrix}
\sigma_1^2 & \rho \sigma_1\sigma_2\\
\rho \sigma_1\sigma_2 & \sigma_2^2
\end{pmatrix},
\label{eq:cov2d}
\end{equation}
where $\sigma_1^2={\rm Var}(\theta_1)$, $\sigma_2^2={\rm Var}(\theta_2)$, and $\rho$ is the correlation coefficient. The corresponding eigenvalues are
\begin{equation}
\lambda_{\pm}
=
\frac{1}{2}
\left[
\sigma_1^2+\sigma_2^2
\pm
\sqrt{(\sigma_1^2-\sigma_2^2)^2+4\rho^2\sigma_1^2\sigma_2^2}
\right].
\label{eq:eigs2d}
\end{equation}
We identify the larger eigenvalue with the direction parallel to the posterior elongation and the smaller one with the perpendicular direction,
\[
\lambda_\parallel\equiv \lambda_+,
\qquad
\lambda_\perp\equiv \lambda_-.
\]
When $|\rho|\to 1$, the posterior becomes highly elongated and
\[
\lambda_\parallel\gg \lambda_\perp,
\]
which is the characteristic signature of a parameter degeneracy.

In the present LISA-only and Taiji-only problem, the broad direction typically corresponds to a family of parameter combinations that produce nearly indistinguishable in-band signals. The posterior extent along that direction is therefore often influenced as much by the prior as by the likelihood. By contrast, the width perpendicular to the ridge measures how tightly the likelihood confines the posterior across the degeneracy. For this reason, the most meaningful single-detector quantity is the perpendicular variance $\lambda_\perp$, rather than the total posterior area or the marginal widths of the original parameters. Indeed, in the Gaussian approximation, the area of a constant-density ellipse scales as
\[
A\propto \sqrt{\lambda_\parallel\lambda_\perp},
\]
so any area-based metric is automatically contaminated by the broad, potentially prior-sensitive direction.

The quantities $\lambda_\parallel$ and $\lambda_\perp$ are variances in log-parameter space, so the corresponding $1\sigma$ widths are $\sqrt{\lambda_\parallel}$ and $\sqrt{\lambda_\perp}$. If
\[
\mathbf e_\perp=(e_1,e_2)^T,
\]
then the $1\sigma$ shifts in the original logarithmic parameters along the perpendicular mode are
\begin{equation}
\delta \theta_1 = \pm \sqrt{\lambda_\perp}\,e_1,
\qquad
\delta \theta_2 = \pm \sqrt{\lambda_\perp}\,e_2.
\end{equation}
Since $\theta_1=\log_{10}\alpha_*$ and $\theta_2=\log_{10}T_*$, these correspond to multiplicative uncertainties in the physical parameters,
\begin{equation}
\alpha_{*,\pm} = \alpha_{*,0}\,10^{\pm \delta\theta_1},
\qquad
T_{*,\pm} = T_{*,0}\,10^{\pm \delta\theta_2}.
\end{equation}
Hence, the upper and lower fractional uncertainties are
\begin{equation}
\Delta \alpha_{*,+} = 10^{\delta\theta_1}-1,
\qquad
\Delta \alpha_{*,-} = 1-10^{-\delta\theta_1},
\end{equation}
and analogously for $T_*$. Since
\[
10^\delta-1 \ge 1-10^{-\delta}
\qquad (\delta>0),
\]
the positive-side fractional uncertainty is always the larger one. It is therefore sufficient to impose a threshold on the positive-side error alone. In particular, requiring this uncertainty to be below $25\%$ gives
\begin{equation}
10^{\sqrt{\lambda_\perp}}-1<0.25
\quad\Longleftrightarrow\quad
\sqrt{\lambda_\perp}<\log_{10}(1.25)\simeq 0.097~{\rm dex},
\end{equation}
or equivalently
\begin{equation}
\log_{10}\!\big(\sqrt{\lambda_\perp}\big)\lesssim -1.01.
\end{equation}
Thus,
\[
\log_{10}(\sqrt{\lambda_\perp})\lesssim -1.01
\]
corresponds to better than $25\%$ reconstruction along the constrained parameter direction.

Given the principal-component decomposition above, parameter-space uncertainties can be propagated into signal-space uncertainties. Let the model signal be denoted by $S(f;\boldsymbol{\theta})$. In the original parameter basis, the usual linear propagation formula is
\begin{equation}
\sigma_S^2(f)
\simeq
\nabla_\theta S(f)^T\,C\,\nabla_\theta S(f),
\qquad
\nabla_\theta S(f)=
\begin{pmatrix}
\partial S/\partial \theta_1\\
\partial S/\partial \theta_2
\end{pmatrix}.
\label{eq:sigprop_orig}
\end{equation}
In the rotated basis defined by the principal components, this becomes
\begin{equation}
\sigma_S^2(f)
\simeq
\left(\frac{\partial S}{\partial q_\parallel}\right)^2\lambda_\parallel
+
\left(\frac{\partial S}{\partial q_\perp}\right)^2\lambda_\perp,
\label{eq:sigprop_rot}
\end{equation}
where $q_\parallel$ and $q_\perp$ denote coordinates along $\mathbf e_\parallel$ and $\mathbf e_\perp$, respectively.

If motion along the ridge changes the observed in-band signal only weakly, the first term in Eq.~\eqref{eq:sigprop_rot} may be neglected to a good approximation. This is the situation relevant to the present domain-wall analysis in the milli-Hz band, where the detector response is dominated by the effective combination associated with the direction perpendicular to the ridge. In that limit, the local signal uncertainty is controlled primarily by $\lambda_\perp$.

For numerical applications, it is often convenient to implement this with finite differences. Let $\boldsymbol{\theta}_0$ denote a reference point, for instance, the posterior mean, MAP point, or injected value, and define
\begin{equation}
\boldsymbol{\theta}_\pm
=
\boldsymbol{\theta}_0 \pm \sqrt{\lambda_\perp}\,\mathbf e_\perp.
\end{equation}
The corresponding model signals are
\[
S_\pm(f)=S(f;\boldsymbol{\theta}_\pm).
\]
The perpendicular-mode signal uncertainty may then be approximated as
\begin{equation}
\sigma_S(f)
\simeq
\frac{|S_+(f)-S_-(f)|}{2},
\label{eq:sigfd}
\end{equation}
with associated fractional uncertainty
\begin{equation}
\frac{\sigma_S(f)}{S(f;\boldsymbol{\theta}_0)}
\simeq
\frac{|S_+(f)-S_-(f)|}{2\,S(f;\boldsymbol{\theta}_0)}.
\label{eq:fracsigfd}
\end{equation}
These quantities should be interpreted as local signal variations induced by the parameter combination that is actually constrained by the data. In the present degenerate problem, they provide the most meaningful notion of signal-side recoverability, precisely because they exclude the broad, prior-sensitive direction that does not correspond to a robust likelihood-driven constraint.

The same principal-component description is also useful beyond the single-detector case. Once the posterior is expressed in terms of its principal variances, degeneracy can be quantified directly through the eigenvalue hierarchy or, in two dimensions, through the axis ratio
\begin{equation}
\kappa \equiv \frac{\lambda_\parallel}{\lambda_\perp}.
\end{equation}
A strongly degenerate posterior is characterized by $\kappa\gg 1$, whereas a more isotropic posterior has $\kappa$ closer to unity. This makes PCA a natural tool for tracking how much the multiband combination reduces the extreme single-detector degeneracy in the joint PTA+LISA and PTA+Taiji analyses discussed below.
\section{Results for LISA(Taiji)-only inference}
\label{sec:singleband}
\begin{figure}
    \centering
\includegraphics[width=.52\linewidth]{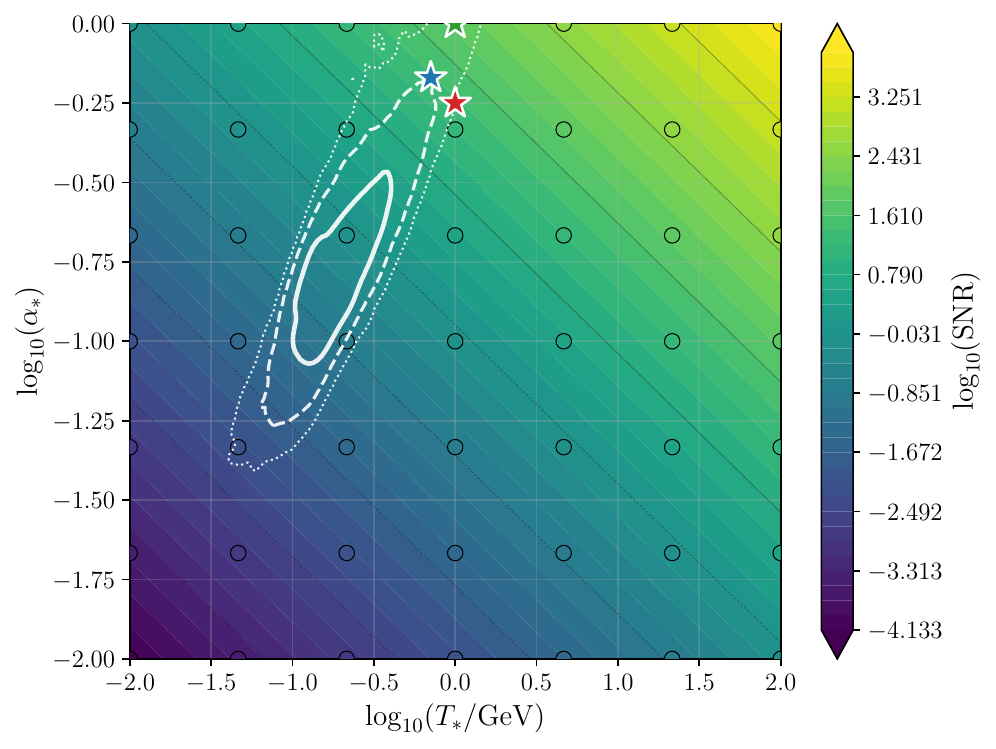}
\includegraphics[width=.47\linewidth]{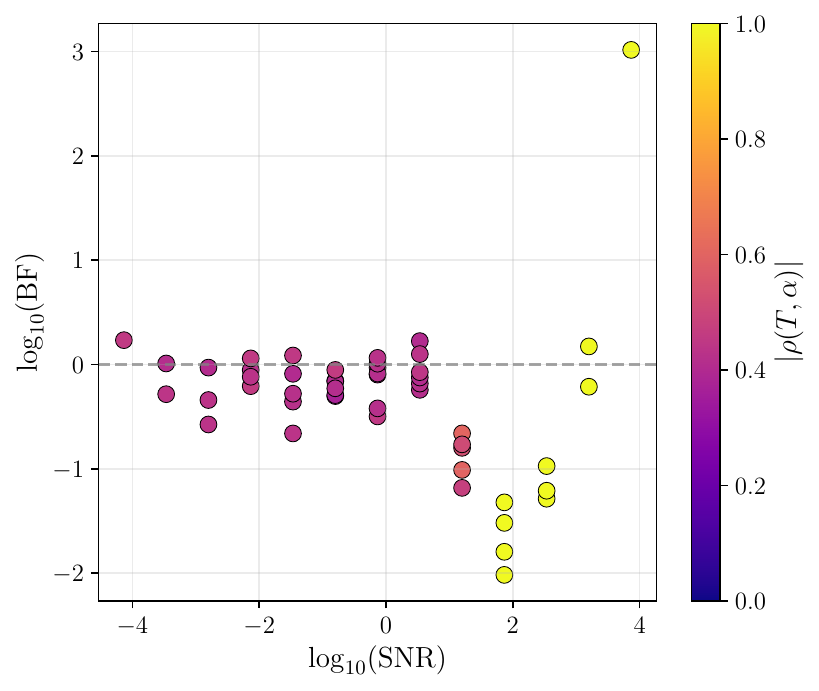}\\
\includegraphics[width=.49\linewidth]{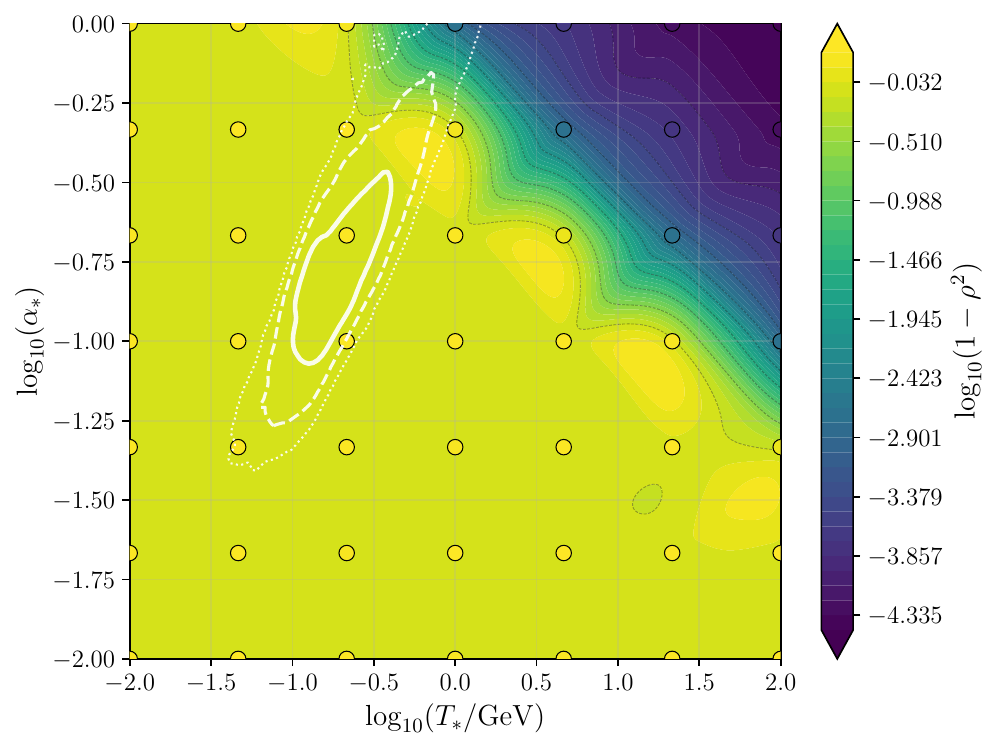}
\includegraphics[width=.49\linewidth]{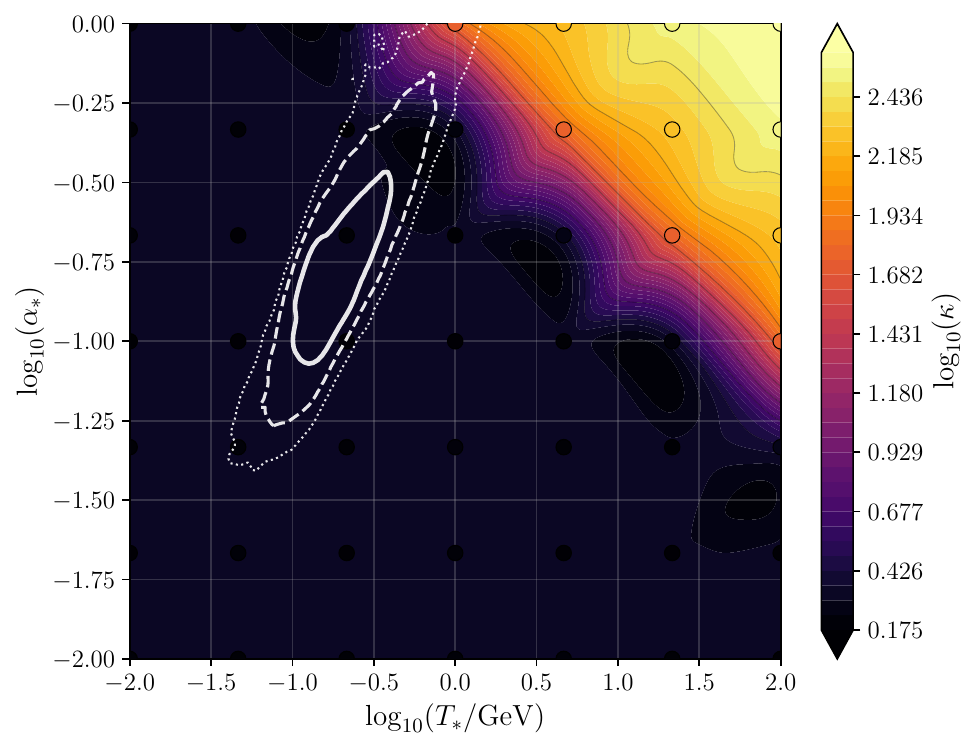}\\
\includegraphics[width=.52\linewidth]{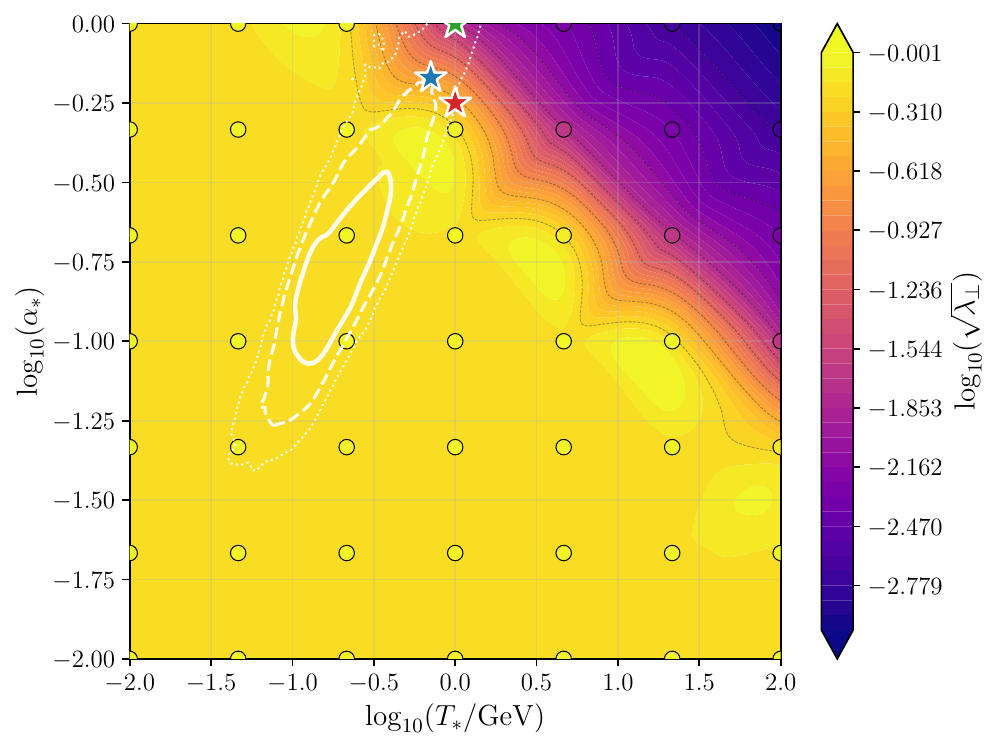}
\includegraphics[width=.47\linewidth]{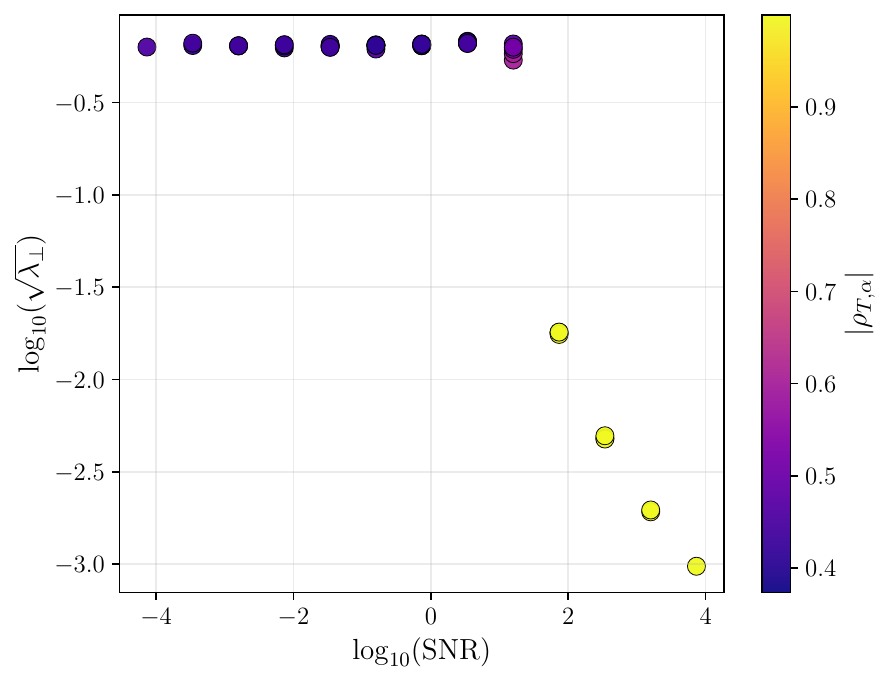}
\caption{Upper panel: Left, signal-to-noise ratio (SNR) on the $\alpha_*$--$T_*$ plane. White contours show the PTA $1\sigma$, $2\sigma$, and $3\sigma$ regions. Colored stars mark three representative injections whose LISA reconstruction prospects are shown in the upper panel of Fig.~\ref{fig:fig3}. Right, $\log_{10}(\mathrm{BF})$ versus SNR. The low-SNR flattening signals the onset of a prior-dominated regime. Middle panel: Left, heat map of $\log_{10}\kappa=\log_{10}(\lambda_\parallel/\lambda_\perp)$ on the $\alpha_*$--$T_*$ plane. Right, heat map of $\log_{10}(1-\rho^2)$ on the same plane, obtained by interpolating 49 equidistant grid injections. Both quantify posterior degeneracy and saturate at low SNR as the inference becomes prior dominated. Lower panel: Left, heat map of $\lambda_\perp$ on the $\alpha_*$--$T_*$ plane. Right, $\lambda_\perp$ versus SNR. The low-SNR plateau again indicates prior-dominated inference. The corresponding LISA results are shown in Fig.~\ref{fig:fig2}.}
    \label{fig:fig1}
\end{figure}
\begin{figure}
    \centering
    \includegraphics[width=.47\linewidth]{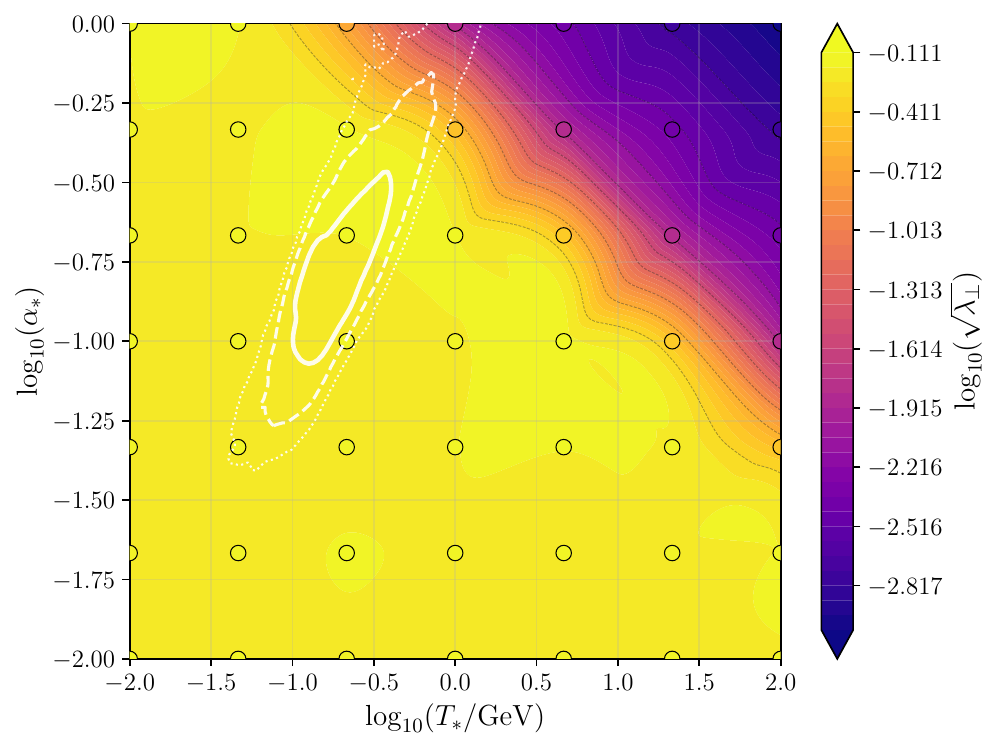}
\includegraphics[width=.47\linewidth]{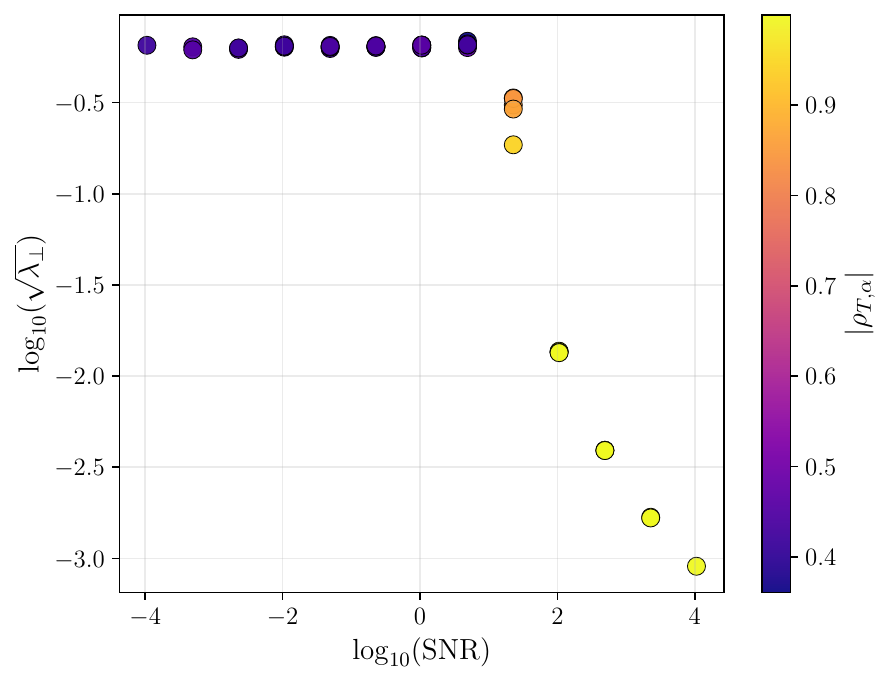}
   \caption{Same as the lower panel of Fig.~\ref{fig:fig1}, but for Taiji. The qualitative behavior is very similar to the LISA case, indicating that the two space-based detectors lead to closely analogous single-detector reconstruction prospects. For this reason, in the joint-analysis section we focus on the PTA--LISA case only.}
    \label{fig:fig2}
\end{figure}
\begin{figure}[h]
    \centering
    \includegraphics[width=.75\linewidth]{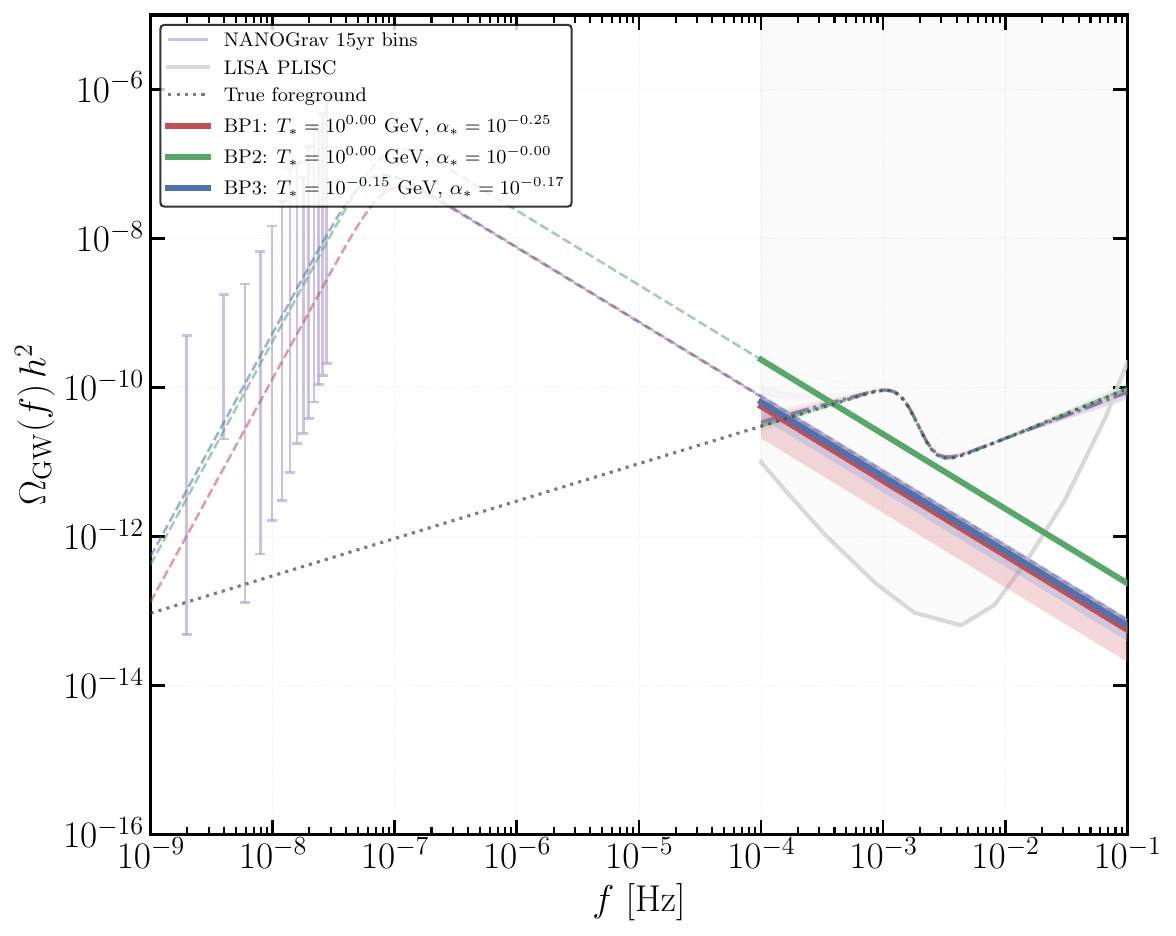}
    \includegraphics[width=.75\linewidth]{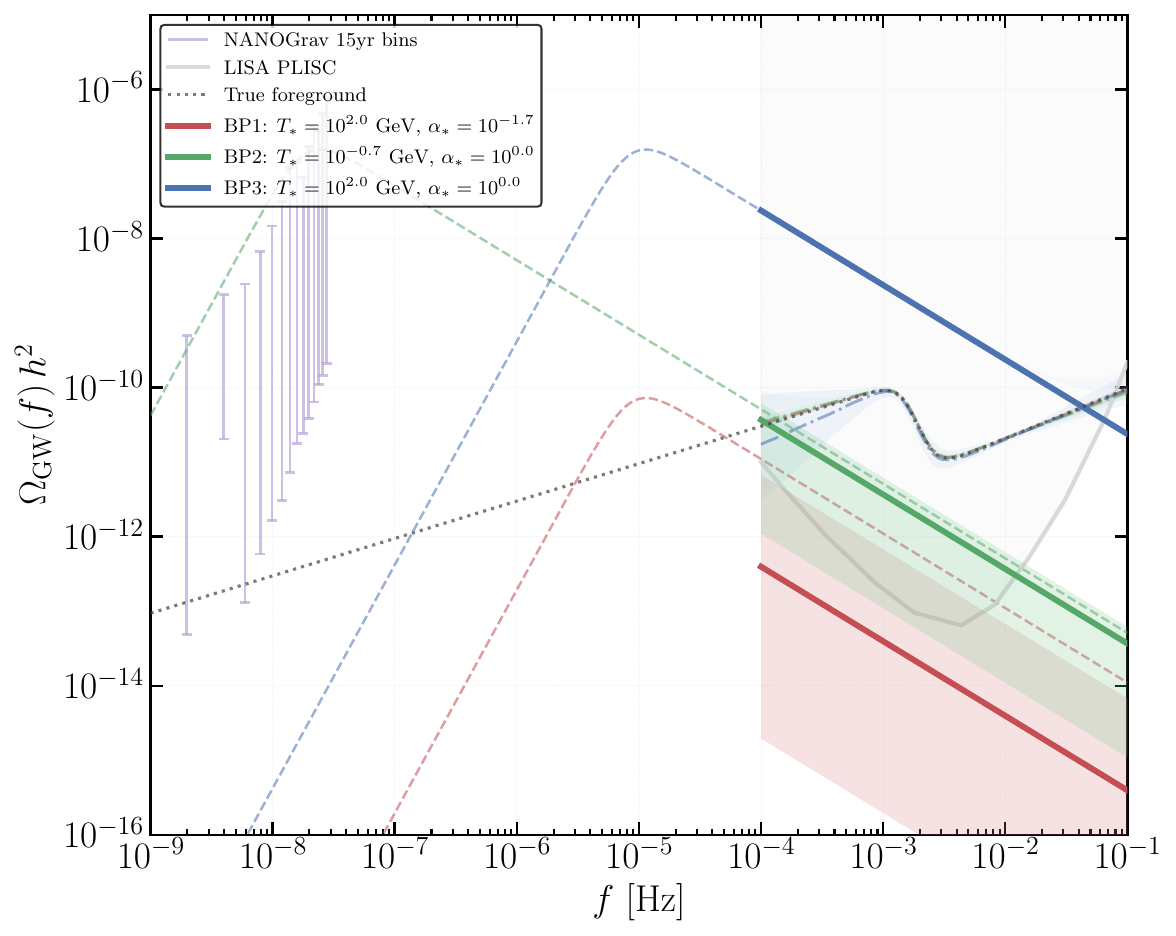}

    \caption{Reconstructed signals at LISA. Upper panel: benchmark signals lying within the PTA-supported region, corresponding to the marked points in Fig.~\ref{fig:fig1}. Lower panel: benchmark signals that produce an ultraviolet tail in the LISA band but lie outside the PTA-supported contours.}
    \label{fig:fig3}
\end{figure}
\begin{figure}
    \centering
    \includegraphics[width=1.1\linewidth]{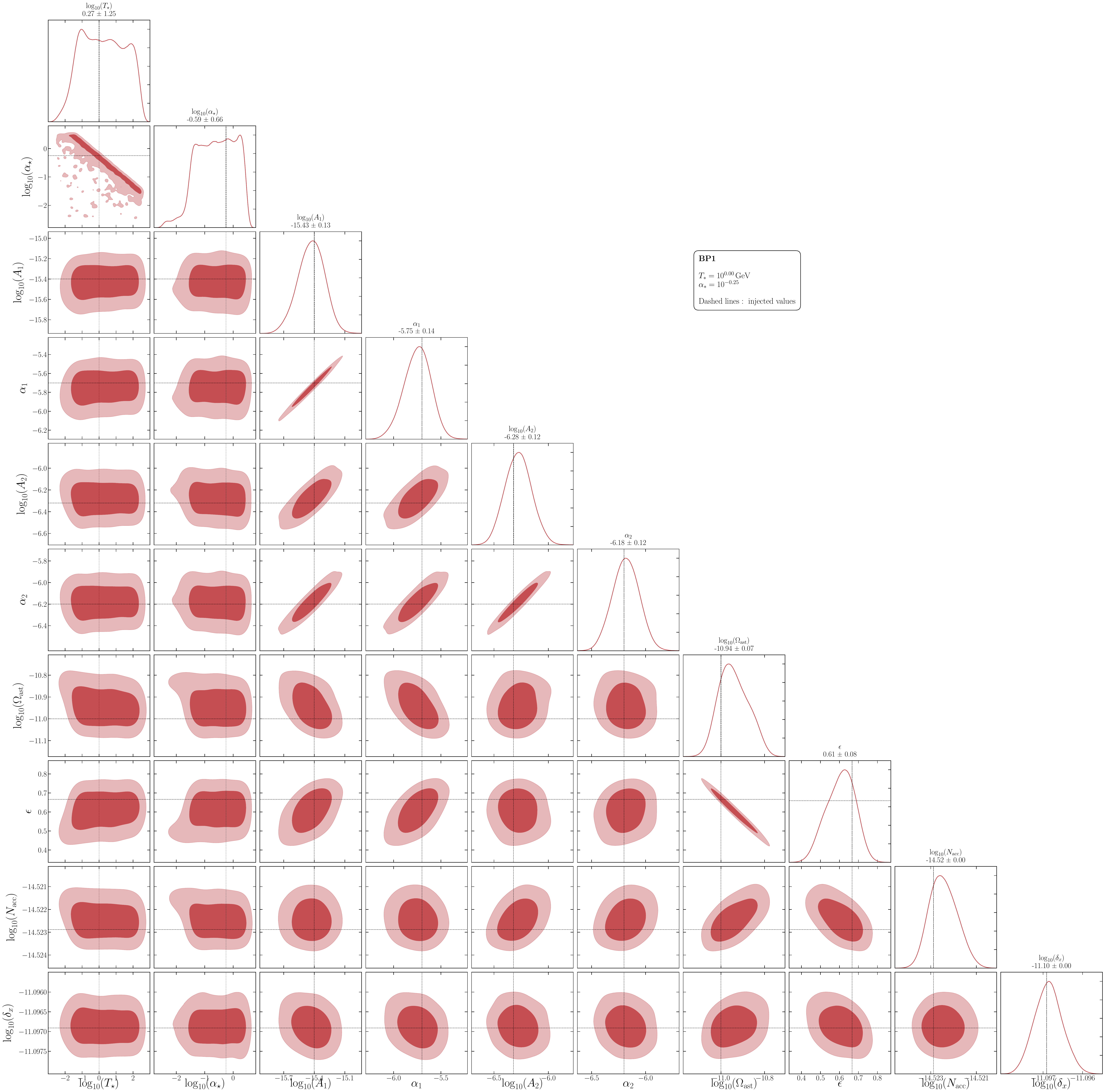}
    \caption{Corner plot for BP1, marked in Fig.~\ref{fig:fig1}. This benchmark lies within the PTA-supported region and illustrates a representative LISA-only reconstruction of an ultraviolet-tail signal, including the two signal parameters and the eight nuisance parameters describing instrumental noise and astrophysical foregrounds. The posterior is strongly degenerate in the signal sector, as expected from the UV-tail nature of the single-detector constraint.}
    \label{fig:fig4}
\end{figure}
\begin{figure}
    \centering
    \includegraphics[width=1.1\linewidth]{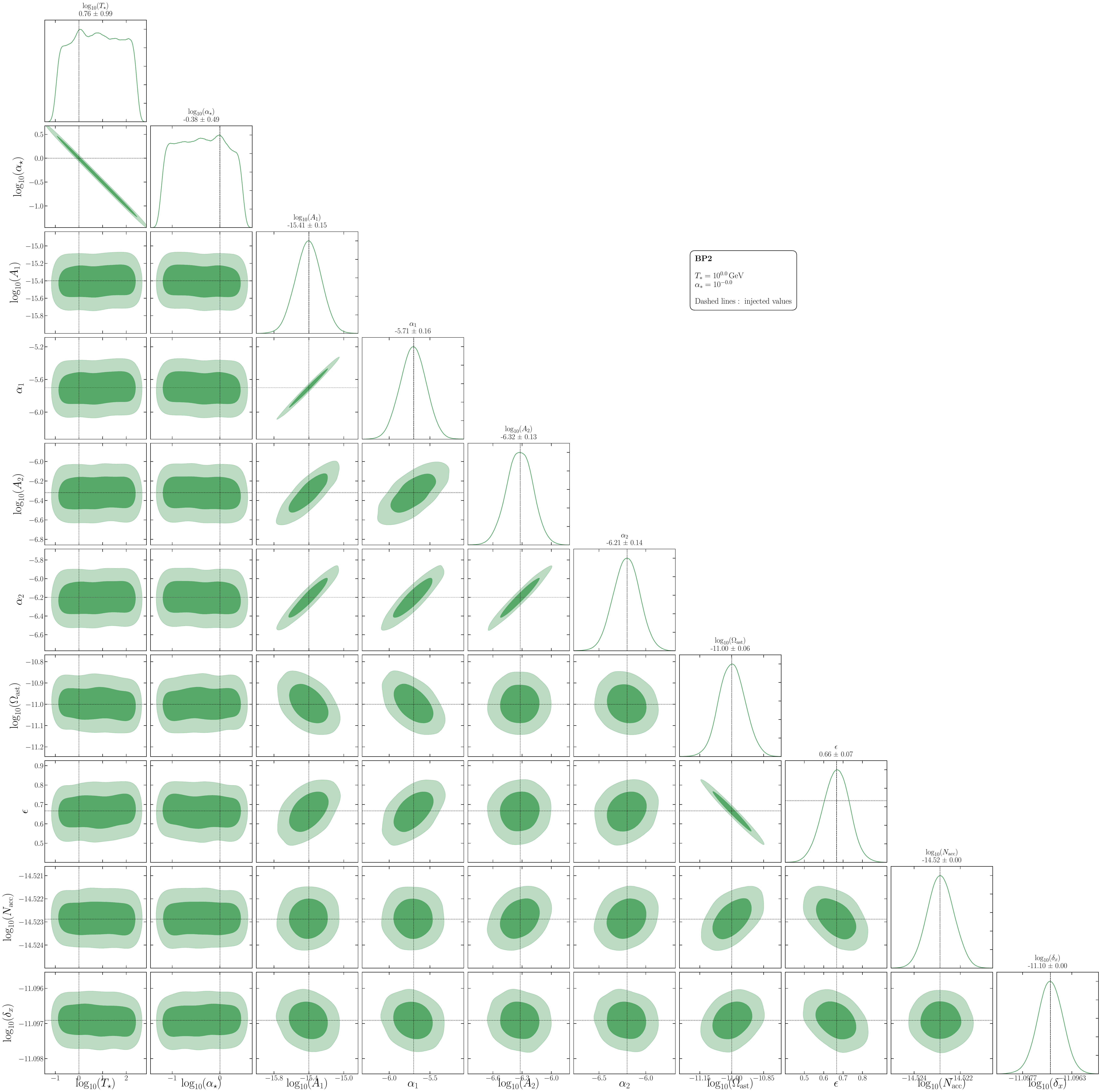}
   \caption{Corner plot for BP2, marked in Fig.~\ref{fig:fig1}. This benchmark lies within the PTA-supported region and illustrates a representative LISA-only reconstruction of an ultraviolet-tail signal, including the two signal parameters and the eight nuisance parameters describing instrumental noise and astrophysical foregrounds. The posterior is strongly degenerate in the signal sector, as expected from the UV-tail nature of the single-detector constraint.}
    \label{fig:fig5}
\end{figure}
\begin{figure}
    \centering
    \includegraphics[width=1.1\linewidth]{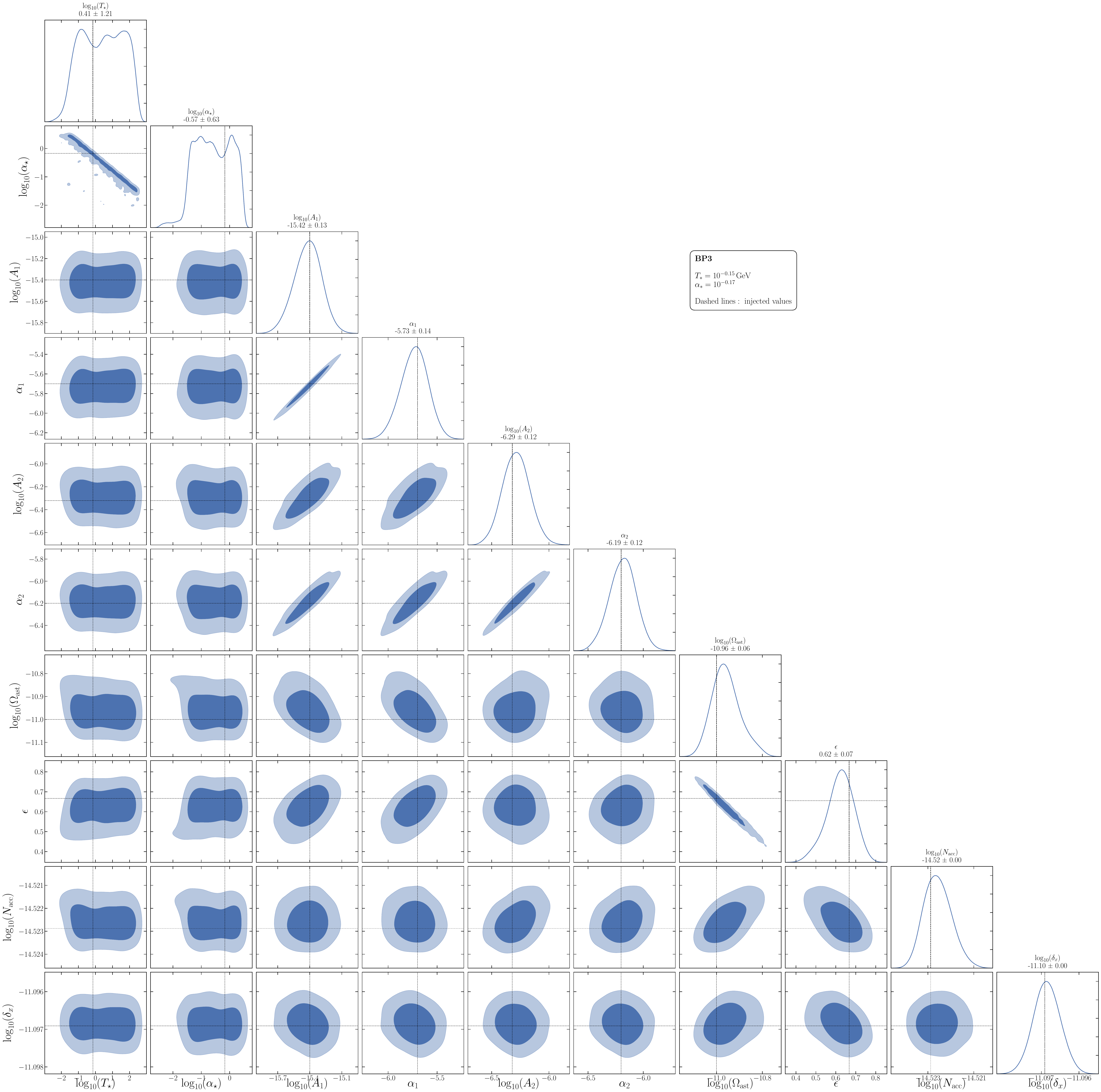}
   \caption{Corner plot for BP3, marked in Fig.~\ref{fig:fig1}. This benchmark lies within the PTA-supported region and illustrates a representative LISA-only reconstruction of an ultraviolet-tail signal, including the two signal parameters and the eight nuisance parameters describing instrumental noise and astrophysical foregrounds. The posterior is strongly degenerate in the signal sector, as expected from the UV-tail nature of the single-detector constraint.}
    \label{fig:fig6}
\end{figure}
We begin with the reconstruction problem based solely on the simulated LISA or Taiji data, using the broad uniform priors for the signal parameters given in Table~\ref{tab:model_parameters} and without imposing PTA posterior information. The aim of this single-detector analysis is twofold. First, it identifies which parts of the domain-wall parameter space are observable once instrumental noise and astrophysical foregrounds are included. Second, it clarifies which combinations of parameters are genuinely constrained by the space-based data alone, and how this constraining power deteriorates as the signal weakens.

Figure~\ref{fig:fig1} summarizes the main results for LISA. The upper-left panel shows the signal-to-noise ratio (SNR) across the $(\alpha_*,T_*)$ plane, with the PTA $1\sigma$, $2\sigma$, and $3\sigma$ contours superimposed in white. This identifies the part of parameter space that yields an ultraviolet tail in the LISA band and, in particular, shows which part of the PTA-supported region remains observable in the presence of instrumental noise and astrophysical foreground contamination. The colored stars mark three representative injection points selected for a more detailed reconstruction study, whose spectra are displayed in the upper panel of Fig.~\ref{fig:fig3}.

The upper-right panel of Fig.~\ref{fig:fig1} displays $\log_{10}(\mathrm{BF})$ as a function of SNR. This provides a model-selection diagnostic complementary to the SNR map, but it should be interpreted with care. The SNR shown here measures signal visibility relative to instrumental noise, whereas the Bayes factor addresses the stricter question of whether the domain-wall component is statistically required once astrophysical foregrounds and their nuisance-parameter freedom are taken into account. Consequently, the two quantities need not turn on at the same threshold. At high SNR, the Bayes factor becomes large and positive, showing that the data favor the inclusion of the domain-wall component over an astrophysical-only model. At intermediate SNR, however, the Bayes factor can remain small or even become negative, despite the fact that the signal is already visible and one parameter combination is beginning to be constrained. This happens because the extended astro+BSM model still contains a broad, weakly constrained direction associated with the signal degeneracy, so that the gain in likelihood is not yet sufficient to overcome the Occam penalty from the enlarged parameter volume. At very low SNR, $\log_{10}(\mathrm{BF})$ approaches a scattered plateau around zero, reflecting the transition to an effectively prior-dominated regime in which the data no longer distinguish the two models in a meaningful way.

The middle panel of Fig.~\ref{fig:fig1} shows two complementary measures of posterior degeneracy across the $(\alpha_*,T_*)$ plane, obtained by interpolating the results from 49 equidistant grid injections. The left plot displays $\log_{10}\kappa=\log_{10}(\lambda_\parallel/\lambda_\perp)$, while the right plot shows $\log_{10}(1-\rho^2)$. Both quantify the degree of elongation of the posterior: large $\kappa$ or, equivalently, small $1-\rho^2$, correspond to strongly degenerate posteriors. These maps make explicit that the LISA-only inference is generically highly anisotropic in the original signal parameters, reflecting the fact that the detector probes only the ultraviolet tail of the spectrum. As the SNR decreases, both degeneracy indicators eventually saturate, again signaling the onset of prior-dominated inference.

The lower panel of Fig.~\ref{fig:fig1} focuses on the quantity most relevant for the single-detector reconstruction problem, namely the perpendicular posterior variance $\lambda_\perp$. As discussed in Sec.~\ref{sec:pca_generalities}, $\lambda_\perp$ measures the width of the genuinely constrained parameter combination, i.e.\ the direction orthogonal to the degeneracy ridge. Unlike the total posterior area, which is contaminated by the broad and often prior-sensitive direction parallel to the ridge, $\lambda_\perp$ isolates the constraining power of the likelihood across the degeneracy. Small values of $\lambda_\perp$ therefore identify regions in which LISA sharply constrains the effective parameter combination governing the ultraviolet tail of the spectrum.

The right-hand plot in the lower panel shows $\sqrt{\lambda_\perp}$ directly as a function of SNR. At low SNR, $\lambda_\perp$ approaches an approximately constant value, indicating that the inference has become prior-dominated and that the data no longer constrain even the perpendicular mode. As the SNR increases beyond this regime, $\lambda_\perp$ drops rapidly, showing that the likelihood begins to resolve the short axis of the posterior and to constrain one principal parameter combination efficiently. This transition can occur already at moderate SNR, before the Bayes factor turns decisively positive. There is no contradiction here: $\lambda_\perp$ is a local reconstruction metric within the extended model, whereas the Bayes factor is a global model-comparison statistic. Thus LISA can begin to measure one signal combination well even in a regime where the astrophysical-only model is still statistically preferred once Occam penalties are included. At very low SNR the Bayes factor scatters around zero, signalling a prior‑dominated regime in which the data cannot distinguish the signal model from the null. At intermediate SNR the Bayes factor can still remain small or even become negative, because the gain in likelihood is not yet sufficient to overcome the penalty for the extra parameter volume—a penalty that is particularly severe when the signal parameters are strongly degenerate. Only for sufficiently strong signals that can break the degeneracy does $\log_{10}(\rm BF)$ turn decisively positive. The existence of high‑SNR points with negative Bayes factors highlights the crucial role of model complexity in gravitational‑wave model selection.

An important point should be emphasized here. The heat maps in Fig.~\ref{fig:fig1} cover a broad $(\alpha_*,T_*)$ region that yields an ultraviolet tail in the LISA band. Over this full plane, one indeed finds an extended region, roughly from the middle toward the upper-right corner, for which $\lambda_\perp$ is small. However, once the PTA-supported region is superimposed, the conclusion becomes more restrictive. Inside the PTA contours, a substantial portion still lies in the saturation regime, where the inference is prior-dominated. In practice, only a fraction of the PTA-supported parameter space, corresponding to the higher-SNR edge of PTA support, exhibits genuinely good single-detector reconstruction prospects. Thus, while LISA can constrain the UV-tail combination well over a broad part of its own parameter plane, within the PTA-compatible region, this happens primarily in the outer, high-signal part of the allowed region.

The corresponding Taiji results are shown in Fig.~\ref{fig:fig2}. The qualitative picture is the same as for LISA. Taiji probes a portion of the PTA-compatible parameter space, the posterior remains strongly degenerate in the original signal parameters, and $\lambda_\perp$ again provides a more faithful characterization of the constraining power than the full posterior area. The detailed numerical values differ because LISA and Taiji have different arm lengths and instrumental-noise parameters, and hence different response functions and sensitivities, but the structure of the inference problem is the same in both cases.

A more direct view of the LISA-only reconstruction is provided in Fig.~\ref{fig:fig3}, where we show six benchmark spectra on the $\Omega_{\rm GW}h^2$--$f$ plane together with the PTA data and the LISA sensitivity curve. The three spectra in the upper panel correspond to the starred injection points in the upper-left panel of Fig.~\ref{fig:fig1}, and thus represent PTA-compatible benchmark signals. The corresponding 10-dimensional corner plots are shown in Figs.~\ref{fig:fig4}, \ref{fig:fig5}, and \ref{fig:fig6}. The lower panel of Fig.~\ref{fig:fig3} displays three additional benchmark signals that do not necessarily fit the PTA data, but nevertheless produce an ultraviolet tail in the milli-Hz band. These examples are included to illustrate more directly how the LISA-only reconstruction behaves as the signal weakens.

In both panels of Fig.~\ref{fig:fig3}, the reconstructed signal width increases as the SNR decreases, as expected from the loss of likelihood constraining power. This trend is especially transparent in the lower panel, which contains two explicitly low-SNR benchmarks. For these cases, the LISA likelihood becomes significantly broader, and in the weakest case the signal width approaches an approximately constant value. This is the signal-space manifestation of the low-SNR plateau seen in $\lambda_\perp$: once the likelihood stops providing substantial information, the reconstructed width is no longer determined by the data and instead approaches a prior-dominated floor.

Taken together, these results establish several important points. First, LISA and Taiji can probe an extended region of the $(\alpha_*,T_*)$ plane that yields an ultraviolet tail in the milli-Hz band, but within the PTA-supported region, the genuinely informative part is more restricted and is concentrated toward the high-SNR edge of PTA support. Second, the resulting posterior is generically not well described in terms of independent constraints on the original physical parameters $(\alpha_*,T_*)$, because the detector band sees only the ultraviolet tail and is therefore sensitive mainly to an effective parameter combination. Third, the perpendicular posterior variance $\lambda_\perp$ provides a robust and physically meaningful measure of what the space-based data actually constrain. Finally, the comparison between $\lambda_\perp$ and the Bayes factor shows that reconstruction and model selection need not occur simultaneously: a single-detector analysis can already constrain one principal signal combination before decisive Bayesian preference for the astro+BSM model is reached.

For this reason, the single-detector analysis should not be interpreted as a claim that LISA or Taiji can fully reconstruct the underlying domain-wall model parameters independently, nor that signal visibility alone guarantees decisive model selection. Rather, it shows that the space-based detectors can measure one specific combination of parameters with high precision over the informative part of the UV-tail region, while leaving the orthogonal combination only weakly constrained. This observation also sets the stage for the multiband analysis. Since the LISA(Taiji)-only constrained mode becomes genuinely informative within the PTA-supported region only toward its high-signal edge, the joint PTA--LISA(Taiji) update will be expected to be most significant precisely in that same region. We now turn to that question in Sec.~\ref{sec:joint}.


\begin{figure}
    \centering
    \includegraphics[width=0.5\linewidth]{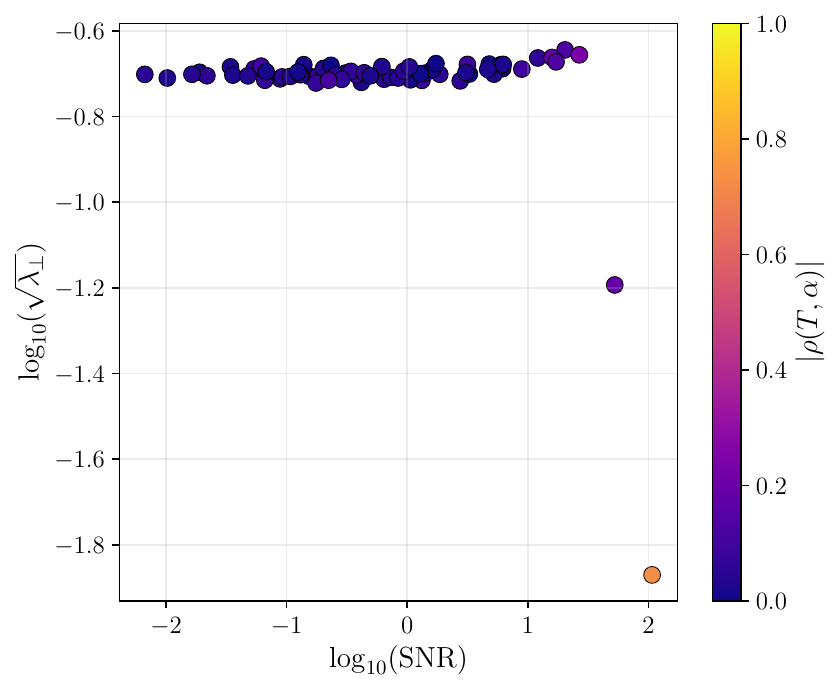}\includegraphics[width=0.53\linewidth]{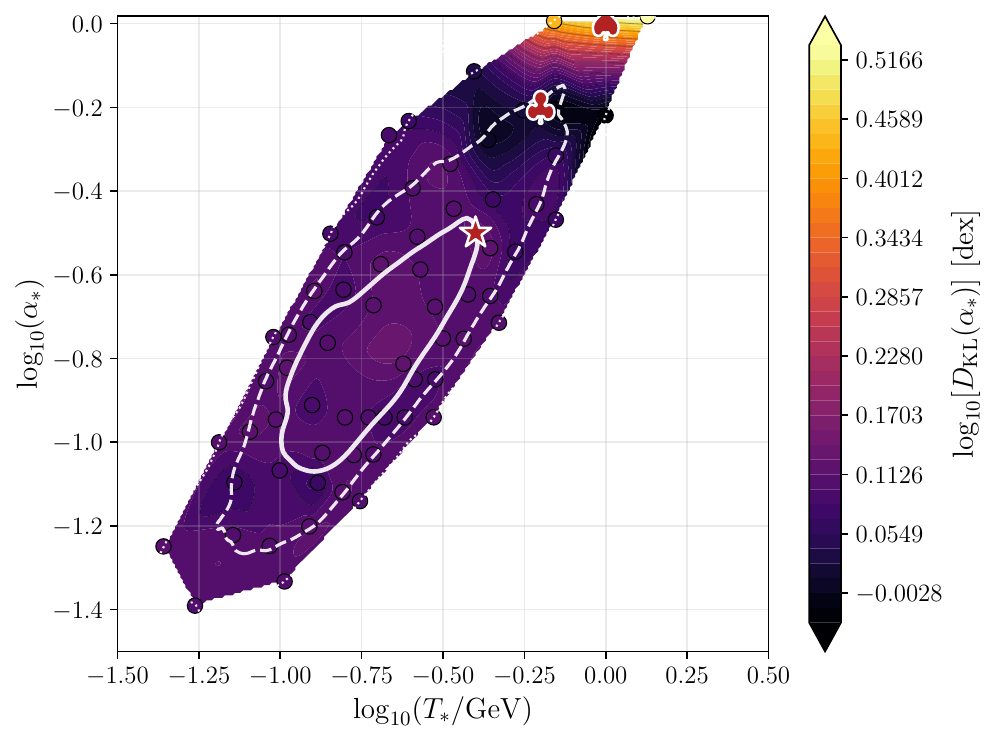}
    \caption{Joint PTA--LISA analysis. Left: $\sqrt{\lambda_\perp}$ as a function of SNR. The trend is qualitatively similar to the LISA-only case: $\lambda_\perp$ saturates in the low-SNR regime, indicating prior-dominated inference, and decreases as the SNR increases, showing that the LISA likelihood progressively constrains the perpendicular mode more efficiently. Right: heat map of $D_{\rm KL}(\alpha_*)$ within the PTA-supported region on the $\alpha_*$--$T_*$ plane; a qualitatively similar map is obtained for $T_*$. The three marked benchmark points are used in Fig.~\ref{fig:fig8}, where the corresponding joint, PTA-only, and LISA-only posteriors are compared directly. The increase of $D_{\rm KL}(\alpha_*)$ toward the high-SNR region shows that LISA can add non-negligible information to the one-dimensional marginal, which is particularly evident for BP2 in Fig.~\ref{fig:fig8}.}
  
    \label{fig:fig7}
\end{figure}
\begin{figure}
    \centering
\includegraphics[width=0.48\linewidth]{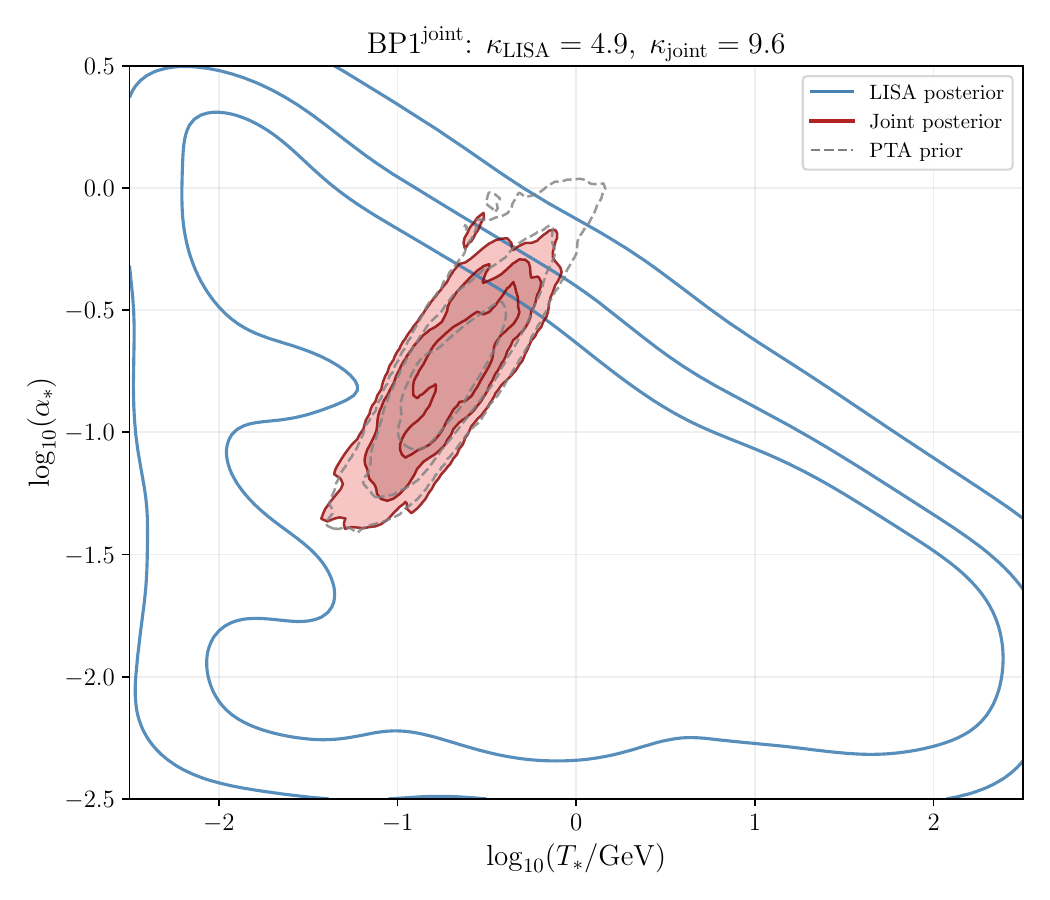}
\includegraphics[width=0.48\linewidth]{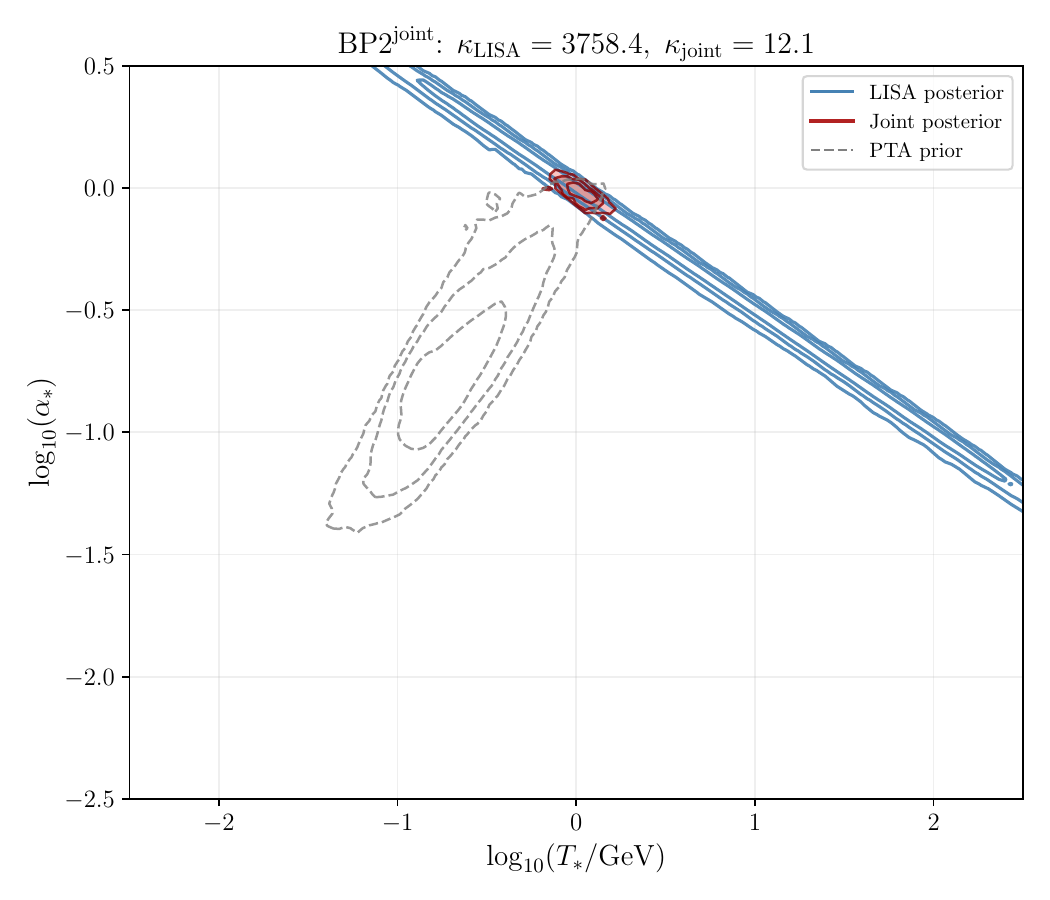}
\includegraphics[width=0.48\linewidth]{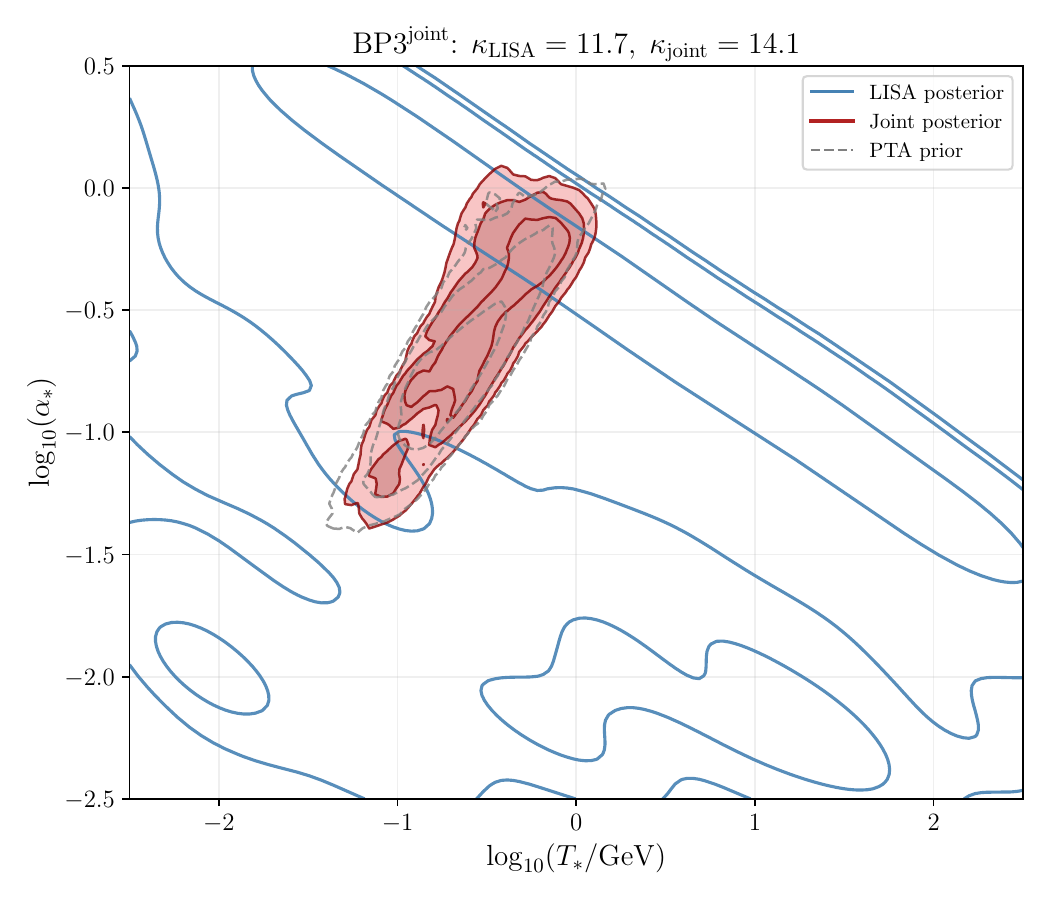}
 \caption{Joint posterior comparison for the three benchmark points shown in Fig.~\ref{fig:fig7}: BP1 (top-left), BP2 (top-right), and BP3 (bottom). The gray dashed contours denote the PTA posterior, the blue contours denote the LISA-only posterior obtained from the same injections using broad uniform priors, and the orange contours denote the joint PTA--LISA posterior. The top-right panel corresponds to a high-SNR case, in which LISA adds appreciable information and the joint posterior is approximately given by the overlap of the PTA and LISA-only constraints. In this case, the extreme single-detector degeneracy is strongly reduced, with $\kappa_{\rm LISA}\sim 3758$ and $\kappa_{\rm joint}\sim 12$. By contrast, the top-left and bottom panels correspond to lower-SNR cases, for which the reconstruction is largely prior dominated: LISA contributes little additional information, and the joint posterior therefore remains close to the PTA posterior used as the prior in the joint analysis. In these low-SNR panels, the corresponding $\kappa_{\rm LISA}$ values should not be over-interpreted, since they are obtained in a prior-dominated regime where $\kappa$ is no longer a reliable measure of genuinely likelihood-driven degeneracy. The corresponding full 10-dimensional corner plots for BP2 is shown in  Fig.~\ref{fig:fig9}.}
    \label{fig:fig8}
\end{figure}

\begin{figure}
    \centering
    \includegraphics[width=1.1\linewidth]{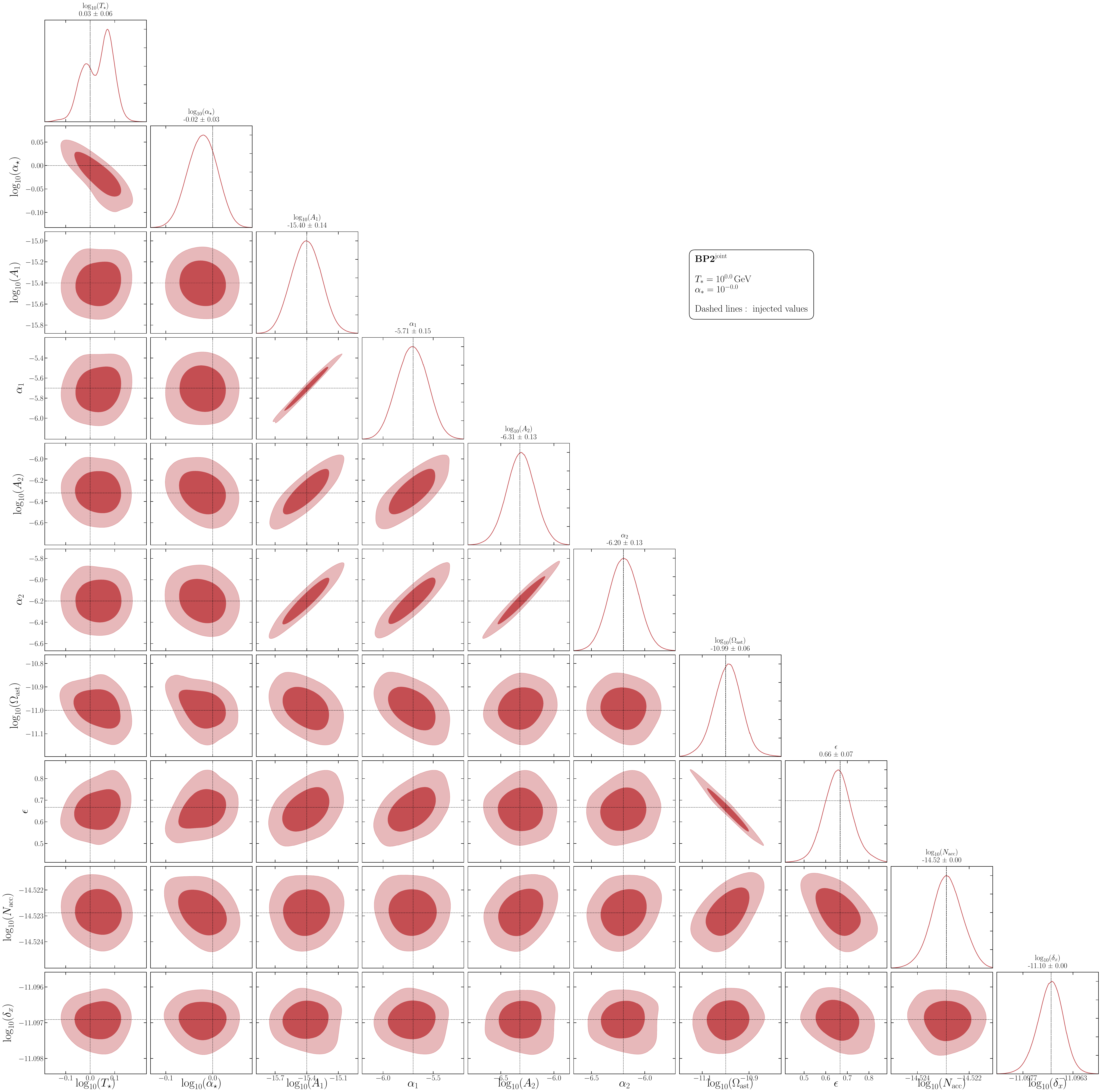}
    \caption{10-dimensional corner plot for BP2, shown in Fig.~\ref{fig:fig8}. This benchmark lies within the PTA-supported region and illustrates a representative joint reconstruction obtained by updating the PTA-informed prior with the LISA likelihood in the informative high-SNR regime.}
    \label{fig:fig9}
\end{figure}

\section{Information gain and degeneracy quantification with joint LISA(Taiji)--PTA analysis}
\label{sec:joint}

We now turn to the multiband inference problem, in which the PTA posterior on the signal parameters is used as an informative prior and subsequently updated with the LISA(Taiji) likelihood.\footnote{We obtain the posterior distribution over the domain‑wall parameters from the NANOGrav 15‑yr pulsar‑timing data using \texttt{PTArcade} \cite{Mitridate:2023oar}. This posterior is then employed as an informative prior for the LISA/Taiji analysis through a sequential Bayesian procedure, where the joint posterior is constructed by importance‑sampling the LISA‑only MCMC samples with the PTA posterior density.} The purpose of this section is twofold: first, to quantify how much additional information the space-based data provide beyond PTA alone; and second, to determine in what sense the joint analysis reduces the strong single-band degeneracy found in the LISA-only case.

A convenient measure of information gain is the Kullback--Leibler divergence (KLD) \cite{Kullback:1951zyt},
\begin{equation}
D_{\rm KL}(p||q)
=
\int dx\, p(x)\,\ln\frac{p(x)}{q(x)},
\end{equation}
where, in the present context, $q$ denotes the PTA posterior used as the prior in the joint analysis, and $p$ denotes the final joint posterior after updating with the LISA(Taiji) likelihood. Thus, $D_{\rm KL}$ measures the average logarithmic update induced by the space-based data. It vanishes when the joint posterior coincides with the PTA posterior and increases as the LISA(Taiji) likelihood shifts and/or sharpens the PTA-supported distribution.

The right panel of Fig.~\ref{fig:fig7} shows the heat map of $D_{\rm KL}(\alpha_*)$ within the PTA-supported region on the $\alpha_*$--$T_*$ plane; a qualitatively similar map is obtained for $T_*$. The main feature is that the information gain is highly localized. It increases toward the outer, high-signal part of the PTA-supported region, where the LISA(Taiji) likelihood becomes strong enough to update the PTA posterior appreciably. By contrast, in the lower-signal part of PTA support, the KLD remains small, indicating that the joint posterior stays close to PTA alone. This is fully consistent with the single-detector results of the previous section: within the PTA-supported region, the perpendicular mode becomes genuinely informative only near the high-SNR edge, and the multiband gain is concentrated in the same region.

The same point is reflected in the left panel of Fig.~\ref{fig:fig7}, which shows $\sqrt{\lambda_\perp}$ as a function of SNR in the joint analysis. The trend is qualitatively similar to the LISA-only case. At low SNR, $\lambda_\perp$ approaches an approximately constant value, indicating that the space-based likelihood is too weak to update the PTA posterior in a meaningful way. As the SNR increases, however, $\lambda_\perp$ decreases, showing that the LISA(Taiji) likelihood begins to constrain the perpendicular mode efficiently. In this sense, the joint analysis becomes informative primarily in the same high-SNR region in which the LISA-only analysis had already indicated nontrivial constraining power on the UV-tail combination.

A direct illustration is provided by the benchmark contours in Fig.~\ref{fig:fig8}, where we compare the PTA posterior, the LISA-only posterior, and the final joint posterior for three representative injection points. In the high-SNR benchmark (BP2, top-right), the joint posterior is localized in the overlap region of the positively correlated PTA posterior and the negatively correlated LISA-only ridge. In this case, the LISA likelihood adds clear information and shifts the posterior toward the injection-consistent overlap region. In the lower-SNR benchmarks, by contrast, the space-based likelihood contributes much less, and the joint posterior remains close to the PTA posterior used as the prior. Thus, the multiband update is substantial only in the region where the LISA-only constrained mode is already informative.

To quantify the posterior geometry, we again use the principal-axis ratio
\begin{equation}
\kappa \equiv \frac{\lambda_\parallel}{\lambda_\perp}.
\end{equation}
As before, large $\kappa$ corresponds to a strongly elongated posterior, while values closer to unity indicate a less degenerate distribution. In the high-SNR region one typically finds
\begin{equation}
\kappa_{\rm joint} \ll \kappa_{\rm LISA},
\end{equation}
showing that the joint analysis efficiently suppresses the catastrophic single-detector degeneracy of the LISA-only posterior. At the same time, one often finds
\begin{equation}
\kappa_{\rm joint} \sim \kappa_{\rm PTA},
\end{equation}
indicating that the joint posterior can retain an axis ratio comparable to that of PTA alone.

This is not contradictory. The quantity $\kappa$ measures the shape of the posterior through the ratio of its principal variances, but it does not encode the absolute size of the allowed region. The joint posterior can therefore be significantly smaller than the PTA posterior while retaining a similar value of $\kappa$. In that situation, the main role of the LISA(Taiji) likelihood is to tighten the PTA-supported region rather than to dramatically reduce its overall elongation.

This point is especially clear in the high-SNR benchmark BP2. There, the LISA-only posterior is extremely elongated, with very large $\kappa_{\rm LISA}$, while the joint posterior has a much smaller axis ratio, showing that the PTA prior efficiently removes the extreme single-band degeneracy. At the same time, the fact that $\kappa_{\rm joint}$ remains of the same order as $\kappa_{\rm PTA}$ shows explicitly that $\kappa$ is insensitive to the absolute size of the posterior. The multiband gain in this regime is therefore best understood not as a dramatic reduction of the PTA elongation, but as a significant tightening of the PTA-supported region together with the removal of the pathological LISA-only ridge.

Taken together, these results lead to a precise interpretation of the multiband gain in the UV-tail regime. The joint analysis is not uniformly informative across the full PTA-supported parameter space. Rather, its impact is concentrated in the outer, high-SNR part of PTA support, where the LISA(Taiji) likelihood can already constrain the perpendicular mode efficiently in the single-detector analysis. In that region, the space-based data produce a non-negligible update of the one-dimensional marginals and substantially reduce the extreme LISA-only degeneracy. In the lower-signal region, by contrast, the joint posterior remains effectively PTA dominated.

\section{Conclusion and outlook}
\label{sec:conclusion}

In this work we have studied the prospects for probing PTA-compatible domain-wall gravitational-wave signals with the milli-Hz space-based interferometers LISA and Taiji, and for quantifying the added value of multiband inference when PTA information is combined with space-based data. The distinctive feature that makes domain walls particularly interesting in this context is the broad ultraviolet tail of their spectrum. Unlike many cosmological GW sources whose signal is effectively confined to a relatively narrow frequency range, domain walls can remain observable in the milli-Hz band even when their peak frequency lies in the nanohertz PTA range. This makes them a natural target for multiband tests.

Our single-detector analysis shows that LISA and Taiji can probe an extended region of the $(\alpha_*,T_*)$ plane that yields an ultraviolet tail in the milli-Hz band, even after instrumental noise and astrophysical foregrounds are included. At the same time, the corresponding single-detector posterior is generically strongly degenerate in the original signal parameters. This is a direct consequence of the fact that LISA and Taiji probe only the ultraviolet tail of the spectrum, so that the data are mainly sensitive to an effective parameter combination rather than to the individual physical parameters separately. For this reason, standard summaries such as marginal widths or posterior areas do not provide the most informative characterization of the constraining power of the experiment.

To address this issue, we developed a principal-component description of the posterior geometry and identified the perpendicular posterior variance $\lambda_\perp$ as the most meaningful local measure of single-detector reconstructability. This quantity isolates the genuinely likelihood-driven direction across the degeneracy ridge and therefore provides a cleaner characterization of the information supplied by the space-based data than the full posterior area. We found that over the full signal plane there is a broad region in which one principal parameter combination can be reconstructed with high precision. However, once the PTA-supported region is superimposed, this conclusion becomes more restrictive: within the PTA contours, a substantial fraction still lies in the prior-dominated regime, and genuinely informative single-detector reconstruction occurs mainly toward the high-SNR edge of PTA support. In this sense, LISA and Taiji can be informative about the domain-wall sector without fully determining the original parameters independently, but that informativeness is localized.

We also investigated Bayesian model selection against an astrophysical-only hypothesis. An important lesson of this analysis is that signal visibility above instrumental noise does not automatically translate into decisive model selection. The instrumental-noise-based SNR and the Bayes factor quantify different notions of sensitivity: the former measures signal visibility relative to detector noise, while the latter tests whether the inclusion of the domain-wall component is statistically warranted once astrophysical foreground freedom and the associated Occam penalty are taken into account. As a result, one may already find small $\lambda_\perp$ and a well-constrained principal mode at moderate SNR while the Bayes factor remains small. This is not an inconsistency, but rather reflects the difference between conditional parameter reconstruction within the extended model and global preference for that model over a simpler astrophysical-only description.

The joint PTA--LISA(Taiji) analysis leads to the central multiband result of this work. Once PTA-informed priors are incorporated, the additional information supplied by the space-based detector is not uniform across the PTA-supported region. Instead, it is concentrated in the same high-signal part of PTA support in which the LISA(Taiji)-only constrained mode had already become informative. In that region, the LISA(Taiji) likelihood produces a non-negligible update of the one-dimensional marginals and substantially suppresses the catastrophic single-detector degeneracy of the LISA(Taiji)-only posterior. At lower signal strengths, by contrast, the joint posterior remains close to the PTA posterior, indicating that the space-based likelihood contributes little beyond PTA alone.

At the same time, our results show that the multiband gain in the UV-tail regime should be interpreted carefully. The joint posterior is often much smaller than the PTA posterior in absolute size, yet its axis ratio can remain comparable to that of PTA alone. Thus, the primary role of the space-based detector is not necessarily to strongly reduce the relative PTA elongation. Rather, its main effect is to tighten the PTA-supported region in the informative high-SNR regime and to remove the extreme single-band LISA(Taiji) degeneracy. Domain walls, therefore, provide a particularly clear example of the nuanced power of multiband GW inference: even when the added information is localized and the relative elongation of the final posterior is not dramatically reduced, combining low- and high-frequency GW observations can still substantially shrink the allowed parameter region and sharpen the interpretation of the inferred source parameters.

More broadly, our results suggest that cosmological GW inference in the multiband era should not be judged solely in terms of standalone detections or large Bayes factors in a single detector. Even when a space-based detector does not by itself provide decisive model selection against an astrophysical-only hypothesis, it may still play a crucial role in tightening the PTA-supported region and improving the physical interpretation of a signal already suggested by PTA data. In this sense, the most important contribution of LISA or Taiji to PTA-motivated domain-wall searches may be geometric rather than purely volumetric.

There are several natural directions for future work. On the signal side, it would be valuable to extend the analysis to a broader class of domain-wall spectral templates and to alternative defect models \cite{Notari:2025kqq,Babichev:2023pbf,Dankovsky:2024ipq}, in order to determine how general the present conclusions are beyond the benchmark template adopted here. On the inference side, a more systematic study of model-selection diagnostics in the presence of strong foreground freedom would help clarify the relation between instrumental-noise detectability, conditional signal reconstruction, and decisive Bayesian evidence for a cosmological component.

A particularly important improvement concerns the modeling of the Galactic foreground. In the present work, we have treated the Galactic double-white-dwarf foreground through an effective isotropic spectral template. This is a natural starting point for forecasting studies, but the true Galactic foreground is known to be anisotropic and time modulated by the orbital motion of the detector \cite{Giampieri:1997ie,Criswell:2024hfn,Digman:2022jmp,Hindmarsh:2024ttn,Buscicchio:2024wwm}. Incorporating this anisotropy in a realistic way could add valuable information in future analyses. In particular, the angular structure and seasonal modulation of the Galactic foreground may help separate it more effectively from an isotropic cosmological component, thereby reducing foreground confusion and potentially strengthening both parameter reconstruction and model selection. A fully anisotropic treatment of the Galactic confusion signal, combined with realistic detector response and time dependence, would therefore be a particularly worthwhile extension of the present study.

It would also be interesting to broaden the multiband framework itself. In addition to PTA and milli-Hz detectors, future decihertz experiments such as DECIGO \cite{Seto:2001qf,Yagi:2011wg} may probe complementary parts of the same cosmological spectrum. Extending the present principal-component and information-gain analysis to such broader detector networks could help clarify how different frequency bands contribute to constraining distinct parameter combinations and which combinations of experiments are most efficient at tightening allowed regions and reducing single-band degeneracies.

In summary, we find that PTA-compatible domain-wall signals can be significantly probed by LISA and Taiji, that the corresponding single-detector posteriors are generically highly degenerate but still informative along one principal direction, and that the combination with PTA information yields a localized but meaningful multiband gain in the high-SNR part of PTA support. Domain walls, therefore, provide a compelling case study for the multiband GW era: they illustrate how the true value of combining datasets may lie not only in improved detectability but also in the tighter and more controlled interpretation that comes from combining complementary constraints across widely separated frequency bands.
\section*{Acknowledgments}
The work of RS is supported by the research project TAsP (Theoretical Astroparticle Physics) funded by the Istituto Nazionale di Fisica Nucleare (INFN).
\appendix
\section{Robustness against alternative spectral-shape parameters}
\label{app:abc_validation}

\begin{figure}
    \centering
    \includegraphics[width=1.1\linewidth]{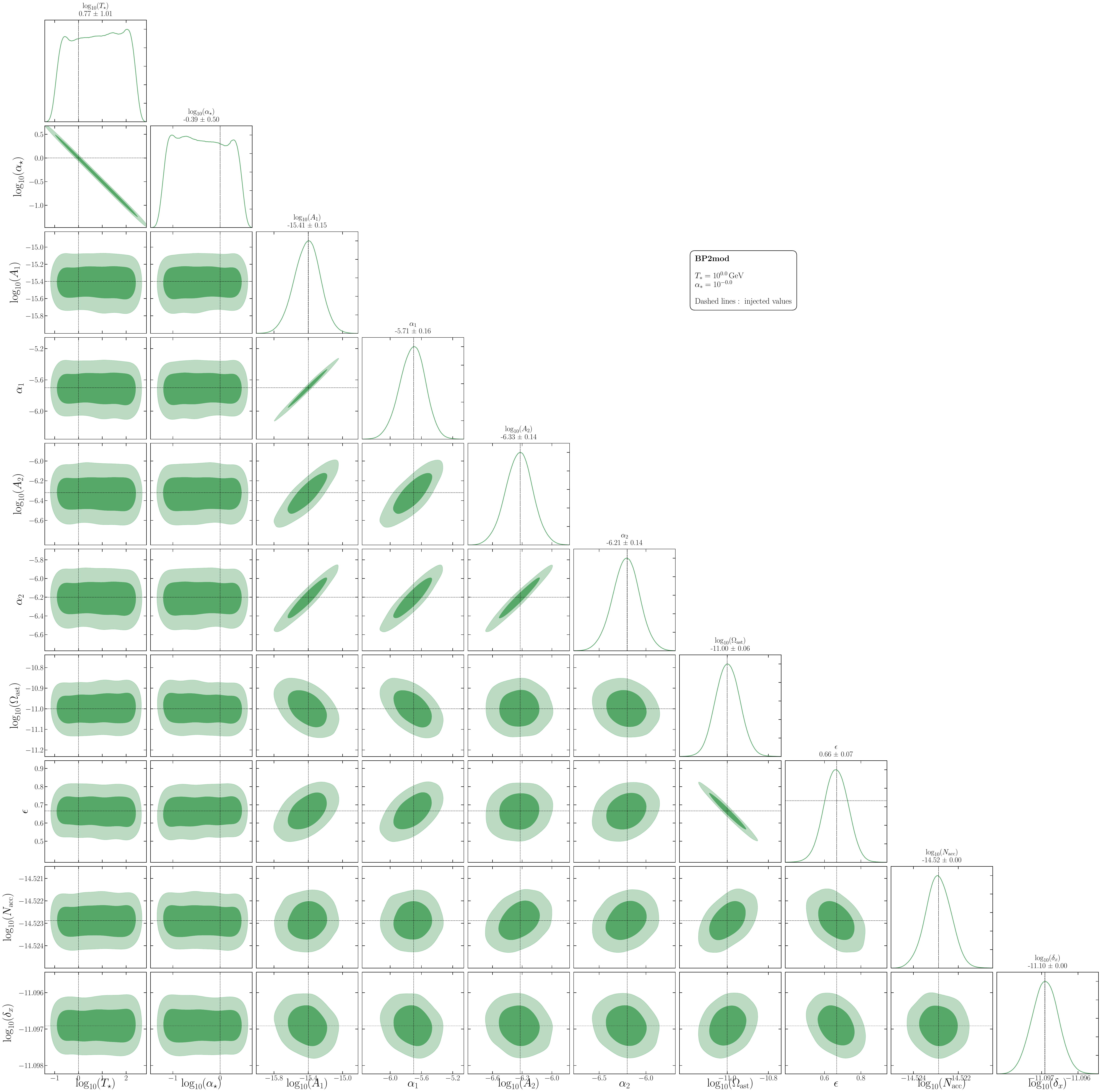}
    \caption{10-dimensional corner plot for the LISA-only reconstruction using the same signal-parameter values as BP2, but with the alternative spectral-shape parameters $a=2.63$, $b=0.86$, and $c=0.86$ motivated by the $2048^3$ lattice simulation results of Ref.~\cite{Babichev:2025stm}. The qualitative posterior structure remains consistent with that shown for BP2 in the main text, indicating that the single-detector reconstruction results are robust against this change in spectral shape.}
    \label{fig:fig10}
\end{figure}

In this appendix we verify that the conclusions of the main text are not sensitive to the specific benchmark choice of spectral-shape parameters,
\[
a=3,\qquad b=1,\qquad c=1.
\]
To this end, we repeat the LISA-only reconstruction for the same signal-parameter values as BP2, but adopt instead
\[
a=2.63,\qquad b=0.86,\qquad c=0.86,
\]
as inferred from a $2048^3$ lattice simulation for a benchmark relative bias in Ref.~\cite{Babichev:2025stm}. The resulting 10-dimensional corner plot is shown in Fig.~\ref{fig:fig10}. As can be seen, the qualitative structure of the posterior remains unchanged: the signal sector still exhibits the same characteristic single-detector degeneracy, and the overall reconstruction behavior is consistent with that found in the main text. This shows that our conclusions are robust against moderate changes in the spectral-shape parameters.

\section{Interpolation method}

To visualize the dependence of the inferred quantities on the injection parameters, we interpolate the discrete forecast results over the sampled parameter plane. Since the injections are not always arranged on a regular lattice, we adopt a triangulation-based interpolation scheme that is applicable to both irregular and equidistant sampling patterns.

We first construct a Delaunay triangulation \cite{LeeSchachter1980} of the injection coordinates. This partitions the sampled region into a set of non-overlapping triangles whose vertices are the injection points themselves \cite{RenkaCline1984}. The Delaunay triangulation is a standard choice because it avoids excessively skinny triangles and provides a numerically stable local tessellation of the sampled domain. It also defines the convex hull of the injection set, which is later used to exclude unsupported regions from the final visualization.

On this triangulation, we construct a Clough--Tocher piecewise-cubic interpolant \cite{Nielson1983}. The resulting surface reproduces the forecast values exactly at all sampled injection points while remaining globally $C^1$ continuous. In practice, this yields smooth contour maps and avoids the sharp gradient discontinuities that would arise from simple linear interpolation within each triangle.

For positive-definite quantities spanning several orders of magnitude, such as $\kappa$ and $\sqrt{\lambda_\perp}$, the interpolation is performed in logarithmic space. Concretely, we interpolate $\log_{10} z$ rather than $z$, which represents any metric itself, and then exponentiate the result. This improves numerical stability and produces more balanced heat maps by preventing the largest values from dominating both the interpolation and the color scale.

The interpolant is then evaluated on a regular mesh covering the sampled parameter region. To avoid extrapolation, all mesh points lying outside the convex hull of the injections are masked. Consequently, the displayed heat maps are smooth visual representations of the sampled forecast surface: they are exact at the injection points and shown only in regions directly supported by the underlying simulations.
\bibliography{bibliography}
\end{document}